\documentclass[12pt]{iopart}
\usepackage{cite}
\usepackage{graphicx}
\usepackage{upgreek}
\usepackage{hyperref}
\usepackage{subcaption}

\usepackage{todonotes}

\newcommand{\MBF}{MBF} % monolithic bonded FIOS
\newcommand{\GMF}{GMF} % Gravity mounted FIOS
\newcommand{\Rx}{Rx} % received beam
\newcommand{\Tx}{Tx} % transmitted beam
\newcommand{\LO}{LO} % local oscillator

\newcommand{\SHS}{SHS}  % Shack-Hartmann sensor 
\newcommand{\refSEPD}{RefSEPD}
\newcommand{\refQPD}{RefQPD}
\newcommand{\tempSEPD}{TempSEPD}
\newcommand{\SciQPD}{SciQPD}

\begin{document}
\title{Reducing tilt-to-length coupling for the LISA test mass interferometer}

\author{M~Tr\"obs$^1$, S~Schuster$^1$, M~Lieser$^1$, M Zwetz$^1$ M~Chwalla$^2$,
	K~Danzmann$^1$, G~Fern\'andez Barr\'anco$^1$,
	E~D~Fitzsimons$^2$\footnote{present address:\ UK Astronomy Technology Centre, Royal Observatory Edinburgh, Blackford Hill, Edinburgh EH9 3HJ, UK},
	O~Gerberding$^1$, G~Heinzel$^1$, C~J~Killow$^1$, M~Perreur-Lloyd$^3$, D~I~Robertson$^3$,
	T~S~Schwarze$^1$, G~Wanner$^1$ and H~Ward$^3$}

\address{$^1$ Max Planck Institute for Gravitational Physics (Albert Einstein Institute) and Institute for
	Gravitational Physics of the Leibniz Universit\"at Hannover, Callinstr.\ 38, 30167 Hannover, Germany}

\address{$^2$ Airbus DS GmbH, Claude-Dornier-Stra{\ss}e, 88090 Immenstaad, Germany}

\address{$^3$ SUPA, Institute for Gravitational Research, University of Glasgow, Glasgow G12 8QQ, Scotland, UK}

\ead{gerhard.heinzel@aei.mpg.de}

\begin{abstract}
	Objects sensed by laser interferometers are usually not stable in position or orientation.
	This angular instability can lead to a coupling of angular tilt to apparent
	longitudinal displacement -- tilt-to-length coupling (TTL).
	In LISA this is a potential noise source for both the test mass interferometer and the long-arm interferometer.
	We have experimentally investigated TTL coupling in a setup representative for the LISA test
	mass interferometer and used this system to
	characterise two different imaging systems (a two-lens design and a four-lens design) both
	designed to minimise TTL coupling.
	We show that both imaging systems meet the LISA requirement of $\pm$25\,$\upmu$m/rad for interfering beams with relative angles of up to $\pm$300\,$\upmu$rad.
	Furthermore, we found a dependency of the TTL coupling on beam properties such as the waist size and location, which we characterised both theoretically and experimentally.
\end{abstract}

\noindent{\it Keywords}: Laser Interferometer Space Antenna, tilt-to-length coupling, test-mass interferometer

% Uncomment for Submitted to journal title message
\submitto{\CQG}

%%%%%%%%%%%%%%%%%%%%%%%%%%%%%%%%%%%%%%%%%%%%%%%%%%%%%%%%%%%%%%%%%%%%%%%
\section{Introduction}

% introduction to LISA, TTL coupling
The space-based gravitational wave detector Laser Interferometer Space Antenna
(LISA)~\cite{elisa13ARXIV,LISAMissionProposal17ARXIV} has been selected as third large-class
mission in ESA's science program~\cite{LISAMission}.
LISA consists of three satellites forming an equilateral triangle with 2.5 million kilometres arm length.
Laser beams are exchanged between satellites and the distance changes caused by gravitational waves
between free-floating test masses inside the satellites are measured.
Telescopes are used for sending and receiving light between spacecraft and the interferometric path length measurements are split in different parts.
Each satellite has optical benches with several interferometers:\ the test mass interferometer measures
distance changes between local test mass and optical bench.
The long-arm interferometer measures distance changes between the local and the remote spacecraft.
To detect gravitational waves these individual measurements are combined to form a Michelson-like interferometer.
The freely-floating test masses and
local interferometry have recently been demonstrated on the LISA Pathfinder (LPF)
spacecraft~\cite{LPF_FEMPTO_G}.

% TTL
Cross-coupling from spacecraft motion into the longitudinal
readout was an important noise source within LPF~\cite{Wanner17JPCS}. This tilt-to-length (TTL) coupling is also a significant contributor in LISA's noise budget.
% other missions, where TTL coupling is important
In addition, TTL coupling is also relevant for the Laser Ranging Interferometer
(LRI) onboard the GRACE Follow-On (GFO) satellites~\cite{Sheard12JGeod}.

% previous work on tilt-to-length coupling
Previously, TTL coupling was reduced in a proof-of-principle experiment by a two-lens imaging
system~\cite{Schuster16OE}.
The reference interferometer in this experiment was made insensitive to TTL coupling~\cite{Schuster15AO} by
using a large single-element photo diode (SEPD) that fully detected the interference of identical Gaussian beams.
In LISA, neither SEPDs can be used nor can the interfering beams be Gaussians with equal parameters.

In LPF the TTL coupling was characterised by intentionally inducing angular variations and measuring the system's response. 
With this knowledge the TTL coupling was subtracted in post-processing~\cite{Wanner17JPCS}. 
For LISA however, this procedure is not sufficient. Due to larger expected angular jitter and more stringed requirements 
additional measures are needed to reduce TTL coupling to a level that can be successfully subtracted in post-processing. 

% TTL coupling in LISA
In LISA, TTL coupling is expected in the test mass interferometer due to spacecraft angular jitter relative to the beam reflected from the 
test mass and in the long-arm interferometer due to angular jitter between the beam received from the far spacecraft and jitter of the local spacecraft.
% this experiment
Hence we built a test bed to experimentally investigate both cases.
The test bed consists of two separate parts:\ an optical bench (OB) and a telescope simulator (TS).
The OB simulates the relevant parts of the LISA optical bench and contains
the measurement interferometer and the imaging systems under test.
The TS is designed to deliver a phase stable, tilting beam (named \Rx{} beam).
The profile of the beam from the TS can be chosen to simulate either the LISA long-arm interferometer
using a flat-top beam, or use a  Gaussian beam to simulate the test mass interferometer.

% goal
The goal of this investigation is to experimentally demonstrate a reduction of TTL coupling in a setup representative for the LISA test mass interferometer to a coupling within $\pm$25\,$\upmu$m/rad for angles
within $\pm$300\,$\upmu$rad between the interfering beams (this complies to $\pm 1\,$pm / $40\,$nrad).
% brief reasoning of requirement:
This requirement comes from a top-level breakdown made in a previous mission study~\cite{Weise10TN6} that was adopted as a conservative and rather stringent requirement.

\begin{figure}[bht]%
	\centering
	\includegraphics[width=0.9\columnwidth]{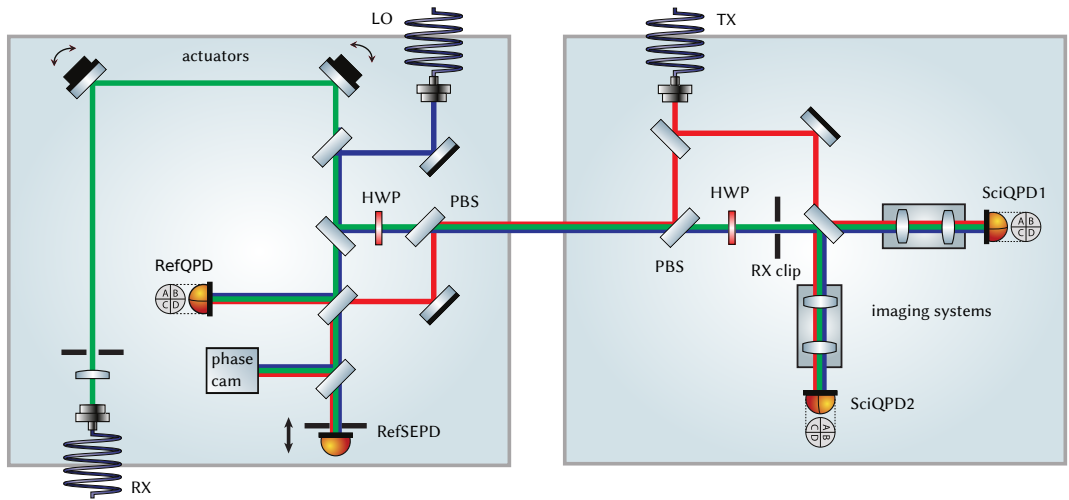}%
	\caption{Schematic of the test bed concept. The Telescope Simulator (left) and Optical Bench (right) are shown with the key components to illustrate the measurement concept. 
		The \Rx{} beam is shown in green, the local oscillator (\LO{}) beam in blue and the \Tx{} beam in red. 
		The \Rx{} beam is tilted around the centre of the \Rx{} clip with the two actuators. 
		The beams between the two baseplates have a different polarization and are separated by polarizing beam splitters (PBS). Source:\ \cite{Chwalla16CQG}}%
	\label{Fig_TestBedConcept}%
\end{figure}
\begin{figure}[bht]
	\centering
	\includegraphics[width=\textwidth,clip=false]{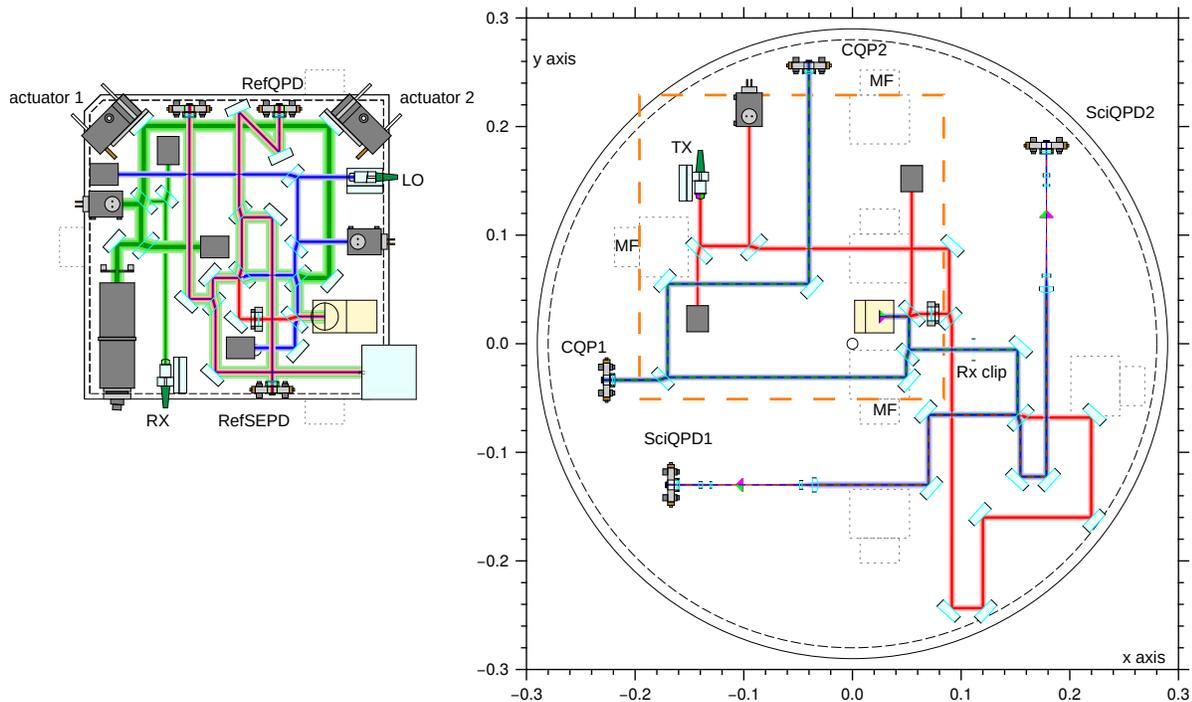}
	\caption{\label{Fig_TSonOB} Optical layout of the test bed. For clarity, the TS is shown next to the OB. 
		In the experiment, it is  placed on top of the OB (shown right) at the position marked with the dashed orange square with dedicated mounting feet (MF). 
		The origin of the coordinate system is on the surface of the OB. The positive z axis points upwards (out of the paper plane). 
		The TS can be adjusted in all three lateral degrees of freedom x, y, z and it can be rotated around all three axes. \Rx{} beam:\ green, \LO{} beam:\ blue, \Tx{} beam:\ red.}
\end{figure}

The test-setup used here is sketched in figure~\ref{Fig_TestBedConcept} and the more complex implementation is illustrated in figure~\ref{Fig_TSonOB}.
A detailed description of the test bed construction, why and how this is representative for LISA, fundamental measurement concepts, as well as the design of the two imaging systems characterised here is given in~\cite{Chwalla16CQG}.

A summary of the measurement concepts used in the test-bed and the principle functions of the two imaging systems tested here, can be found in section~\ref{sec:Measurement_concepts}. 
The preparation and characterisation of the laser beams, as well as alignment procedures and resulting alignment precisions of the test bed and its calibration, are given in section~\ref{sec_Experiments}.  
The measured TTL suppression results and the dependency on beam parameters are presented in section~\ref{Sec_Results}.

%%%%%%%%%%%%%%%%%%%%%%%%%%%%%%%%%%%%%%%%%%%%%%%%%%%%%%%%%%%%%%%%%%%%%%%
\section{Measurement concepts}
\label{sec:Measurement_concepts}

\subsection{Working principle of imaging systems}
\label{sec:IS_concepts}
Two imaging system designs were tested: a four-lens system designed with classical concepts for pupil plane imaging, and a two-lens system which unlike the four-lens system generates a diverging output beam. 
Both imaging system types are designed to suppress the TTL coupling by imaging the point of rotation in the \Rx{} clip to the quadrant photodiodes named \SciQPD{}1 and \SciQPD{}2.
To avoid systematic errors, two identical copies of each imaging system design were manufactured, and one copy placed in front of each \SciQPD{}.
Each measurement was performed with both copies simultaneously in the two output ports. 
Details on the optical and mechanical imaging system designs, requirements, and specifications are given in~\cite{Chwalla16CQG}.

%%%%%%%%%%%%%%%%%%%%%%%%%%%%%%%%%%%%%%%%%%%%%%%%%%%%%%%%%%%%%%%%%%%%%%%
\subsection{Test bed summary}
\label{sec: test bed summary}
%\Ueberlick
In the context of this work, the aim of the test bed is to validate that TTL coupling, 
(i.e. the cross coupling of beam tilt caused by angular misalignment between test mass and optical bench to the interferometric phase) 
in a system representative for the LISA TM interferometer can be characterised and suppressed. 
In order to investigate this TTL coupling, the test bed does not feature a test mass (TM), but only a Gaussian beam (named here \Rx{} beam, 
see figure~\ref{Fig_TestBedConcept}) which rotates around a fixed point (labelled \Rx{} clip) which represents the reflection point on the rotating TM. 
We then measure the TTL coupling between the tilting \Rx{} beam and a fixed Gaussian reference beam named \Tx{} beam on quadrant photodiodes 
(\SciQPD{}1, \SciQPD{}2) at distances behind the rotation point representative for the LISA OB and show that imaging systems reduce this coupling to below the given requirement.

% Erste Feinheit: Rotation um Rx-Clip Teil 1
Our chosen measurement concept requires that the \Rx{} beam rotates around the fixed \Rx{} clip. 
This requires that there is no lateral beam walk in the \Rx{} clip during the beam rotation (lateral alignment of the rotation point), 
and that the phase of the \Rx{} beam relative to the \Tx{} beam in the \Rx{} clip does not change during rotations induced by the actuators on the TS. 
The longitudinal requirement originates from the necessity to remove (or measure) any  tilt dependent path length change that is induced up to the \Rx{} clip. 
Furthermore, lateral beam walk in the \Rx{} clip would lead to additional phase variations which are unwanted here, 
specifically because lateral motion in the rotation plane is negligible in the TM interferometer.   

\begin{figure}
	\centering
	\includegraphics[width=\textwidth,clip=false]{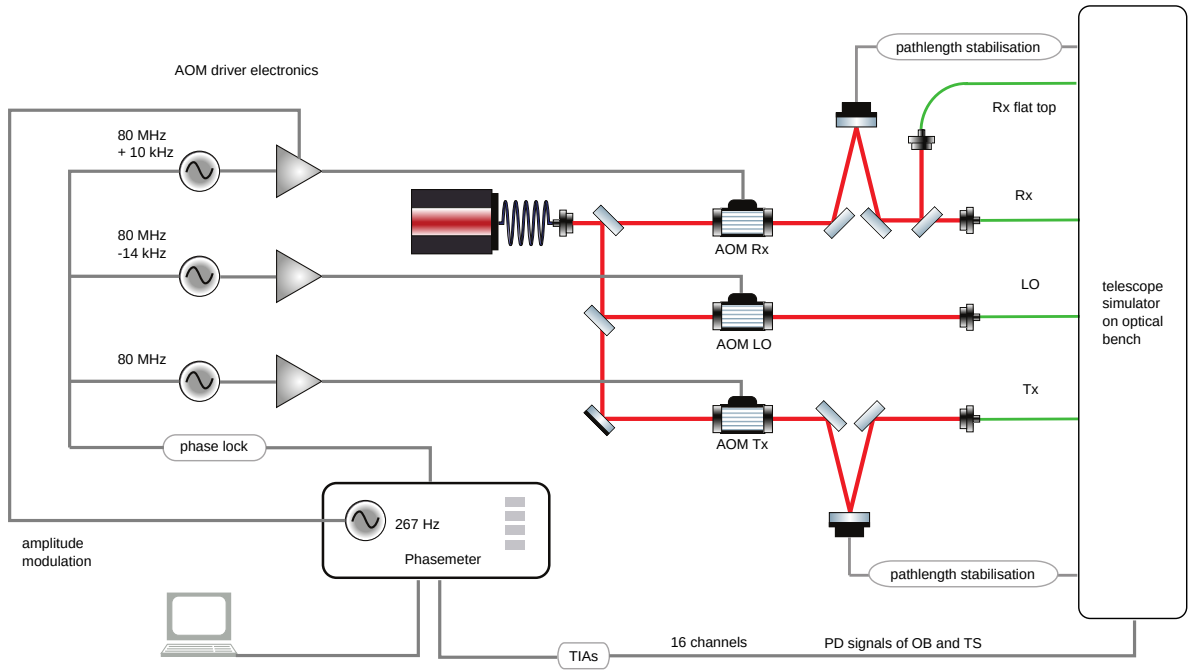}
	\caption{\label{Fig_ModulationBenchElectronics} Schematic of laser preparation, electronics,
		and interferometer readout. Fibres:\ green, laser beams:\ red, cables:\ grey}
\end{figure}

In order to ensure these two requirements are met, the beam walk and the phase difference need to be monitored and controlled in the \Rx{} clip. 
However, it is not possible to measure those quantities directly in the \Rx{} clip without blocking the beam path to the \SciQPD{}s. 
Therefore, a \refSEPD{} and a \refQPD{} are placed at positions equivalent to that of the \Rx{} clip. 
With ``equivalent position'' we mean, that a beam rotated around the centre of the \refQPD{} or \refSEPD{} also rotates around the centre of the \Rx{} clip.
We use the differential power sensing (DPS) signals~\cite{Wanner12OC} of the \refQPD{} to measure lateral displacement of the \Rx{} beam in the \Rx{} clip. 
The beam walk is then suppressed by nulling the DPS signal using actuator~2 (see figure~\ref{Fig_TSonOB}).
The longitudinal alignment of the rotation point is controlled by feeding the \refSEPD{} 
phase signals back to actuators on the modulation bench (path length stabilisation in figure~\ref{Fig_ModulationBenchElectronics}), 
thereby locking the \Tx{} and \Rx{} beam to the local oscillator (\LO{}) beam. %
This minimizes the phase variations between the beams in the location of the \Rx{} clip.
The method of fixing the centre of rotation in the \Rx{} clip by phase locking the \Rx{} and \Tx{} to the \LO{} beam on the \refSEPD{} 
is  achieved by using a 150\,$\upmu$m diameter ``pinhole'' aperture attached to the \refSEPD{}.
A rotation of curved wavefronts around the centre of a larger diode would lead to TTL coupling~\cite{Schuster17CQG}.
The phase lock would then compensate the TTL coupling with a longitudinal shift of the \Rx{} beam, inducing an unintended phase error in the \Rx{} clip.
For a pinhole diode the effective wavefront curvature with respect to the diodes size is negligible, such that the plane wave approximation is valid.
The described effect was validated for the chosen pinhole size and the anticipated wavefronts using the simulation software IfoCAD~\cite{Wanner12OC}.
Therefore, the phase lock on the pinhole \refSEPD{} ensures rotation of the \Rx{} beam around the centre of the \Rx{} clip.

Any residual phase variations that are not cancelled by  this phase lock are subtracted from the \SciQPD{}s signals in post processing.

% Verwendung LO zum Ausrichten von TS vs OB.
Since the test bed is split into the TS and the OB, the described measurement principle naturally requires that these two parts are well aligned with respect to each other. 
This is achieved by aligning the \LO{} beam to a pair of pre-aligned quadrant diodes (CQP1, CQP2 shown in figure~\ref{Fig_TSonOB}). 
As described in \cite{Chwalla16CQG} the \LO{} beam is then  optimally aligned to the \SciQPD{}s  and the TS is optimally aligned to the OB.

%Summary
Details on the describes test bed methodology including alignment and calibration of the various subsystems are given in section~\ref{sec_Experiments}.
For instance, more details on the longitudinal positioning and lateral alignment of the \refSEPD{} and \refQPD{} 
to the equivalent position of the \Rx{} clip is given in section~\ref{Sec_RefIfoLoPos} and section~\ref{Sec_RefIfoAlignment}. Further details on the alignment of the TS to the OB are given in section~\ref{Sec_TSalignment}.

%%%%%%%%%%%%%%%%%%%%%%%%%%%%%%%%%%%%%%%%%%%%%%%%%%%%%%%%%%%%%%%%%%%%%%%
\section{Experiments}
\label{sec_Experiments}
%%%%%%%%%%%%%%%%%%%%%%%%%%%%%%%%%%%%%%%%%%%%%%%%%%%%%%%%%%%%%%%%%%%%%%%
\subsection{Laser preparation and beam characteristics}
\label{Sec_LaserPreparation}
Figure~\ref{Fig_ModulationBenchElectronics} shows the schematic of laser preparation, electronics and the interferometer readout.
The laser is a nonplanar ring oscillator~\cite{Kane85OL, Freitag95OC}.
The beam is divided into three parts to generate the three different frequencies for the OB and TS.
The frequency is shifted by acousto-optical modulators (AOMs), the frequency shifts and resulting heterodyne frequencies are
listed in table~\ref{Tab_AOMfreq} and table~\ref{Tab_hetfreq}.
\begin{table}
	\caption{AOM and heterodyne frequencies}
	\begin{subtable}[c]{0.4\textwidth}
		\centering
		\subcaption{AOM driving frequencies}
		\label{Tab_AOMfreq}
		\begin{tabular}{@{}c l@{}}
			\br
			beam  & frequency (MHz) \\
			\hline
			\LO{} & 79.9853515625   \\
			\Tx{} & 80.0            \\
			\Rx{} & 80.009765625    \\
			\br
		\end{tabular}
	\end{subtable}%
	\begin{subtable}[c]{0.6\textwidth}
		\centering
		\subcaption{Heterodyne frequencies}
		\label{Tab_hetfreq}
		\begin{tabular}{@{}c c r@{}}
			\br
			phase & beams       & frequency (kHz) \\
			\mr
			A     & \Tx{}-\Rx{} & 9.7656250       \\
			B     & \Tx{}-\LO{} & 14.6484375      \\
			C     & \Rx{}-\LO{} & 24.4140625      \\
			\br
		\end{tabular}
	\end{subtable}%
\end{table}
The different heterodyne signals are called A, B and C in the following.
They were chosen using the following criteria:
\begin{itemize}
	\item They shall be of the form $k\cdot f_{\rm PM}/2^{24}$ with $k$ integer and the sampling frequency $f_{\rm PM}=80$\,MHz of the phasemeter.
	      This ensures that the heterodyne frequencies can be generated within the phasemeter.
	\item Heterodyne frequencies B and C should be as high as possible to enable high bandwidth control loops.
	\item Heterodyne frequencies in the range of 10-30\,kHz seemed to be sufficient.
	\item Heterodyne frequencies must not be harmonics of each other.
	\item Heterodyne frequencies must not be harmonics of the power grid frequency of 50\,Hz.
\end{itemize}
The three beams are delivered to the setup by optical fibres. 
Before the delivery to the test bed, the \Rx{} beam is split again and coupled into
two different fibres to allow an easy switching between a Gaussian and a flat-top shaped \Rx{} beam.
However, this work concentrates on the test mass interferometer, so the flat-top option is not used in the scope of this work.

In the beam path of the \Tx{} and the \Rx{} beams there are additional linear piezo actuators after the AOMs.
They are used for optical pathlength stabilisations  (phase lock between \Tx{}, \Rx{} and \LO{}) implemented in the phasemeter, as previously described in section~\ref{sec: test bed summary}.

\subsection{Beams}
\label{sec:Beams}
The beam parameter dependency of the TTL coupling, which will be described in detail in section~\ref{Sec_Results} and~\ref{Sec_Simulations}, generated a need to alter the \Rx{} beam parameters. 
Therefore, we distinguish between the \Rx{} beam delivered from the monolithically-bonded \Rx{} 
fibre output coupler (labelled \MBF{}) and the alternative \Rx{} beam delivered by a gravity-mounted fibre output coupler (named \GMF{}).   

\begin{figure}
	\centering
	\frame{\includegraphics[width=0.75\textwidth]{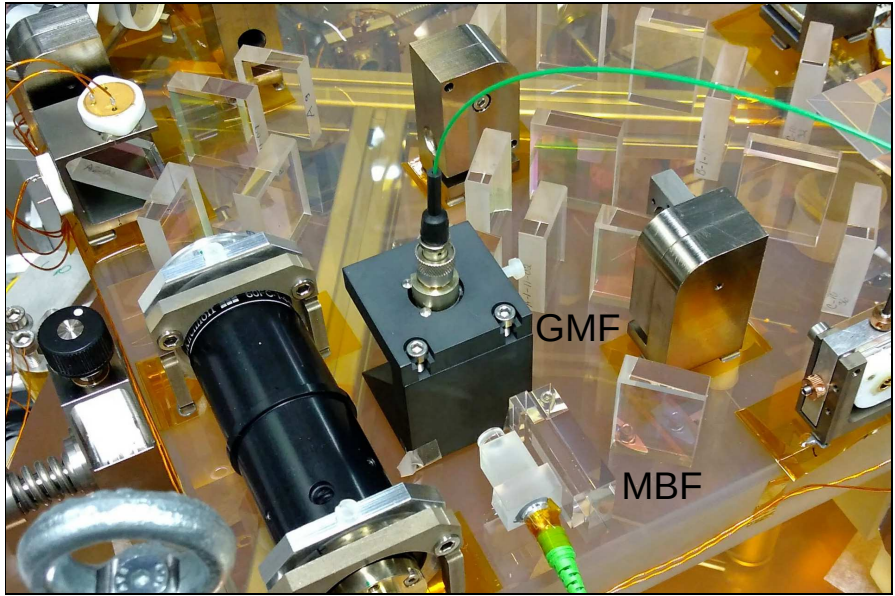}}
	\caption{\label{Fig_GMF-on-TS}\Rx{} beam delivery to the telescope simulator. 
		It is shown how the \MBF{} beam can be exchanged by the \GMF{} beam. 
	}
\end{figure}

The \MBF{}, silicate-bonded~\cite{HC_Bonding2014} to the TS, originally delivered the \Rx{} beam. 
The \GMF{} was designed to temporarily block the \MBF{} and thereby easily alter the beam parameters of the \Rx{} beam as shown in figure~\ref{Fig_GMF-on-TS}.  
The \GMF{} consists of a commercially available fibre coupler (60FC-4-A15-03 by Sch\"after+Kirchhoff) mounted in a
brass mount with a folding mirror.
% MT: added sentence below
The \GMF{} could be placed and aligned ``by hand'' on the TS because the two actuators could be used for the fine alignment of the \Rx{} beam.

\begin{figure}
	\centering
	\includegraphics[angle=-90,width=0.75\textwidth]{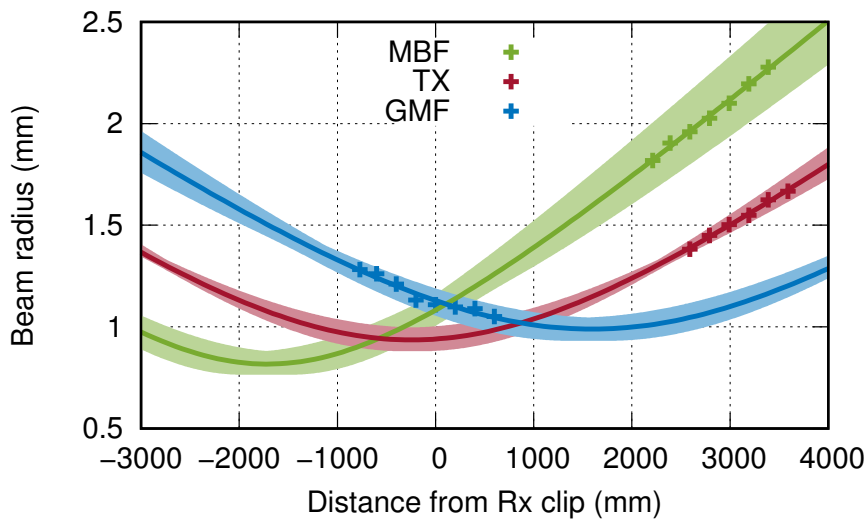}
	\caption{\label{Fig_beams} Beam radii for \MBF{}, \Tx{} and \GMF{} beams as function of distance to the \Rx{} clip; 
		plus signs indicate Gaussian beam radii obtained from fitting Gaussian intensity profiles to measured intensity distributions, 
		lines indicate equation~(\ref{Eq_wofz}) fitted to the measured beam radii with fit parameters $w_0$ and $z_0$. 
		The shaded areas are the 95\% confidence intervals for beam radii resulting from the respective confidence intervals for $w_0$ and $z_0$. 
		Table~\ref{Tab_beams} summarises the beam parameter values plotted here.}
\end{figure}

Figure~\ref{Fig_beams} shows beam radii for \MBF{}, \Tx{} and GMF beams as function of distance to
the \Rx{} clip.
Plus signs indicate Gaussian beam radii obtained from fitting Gaussian intensity profiles to measured intensity distributions.
Lines indicate equation~(\ref{Eq_wofz}) fitted to the measured beam radii with fit parameters $w_0$ and $z_0$.
The shaded areas are the 95\% confidence intervals for beam radii resulting from the respective
confidence intervals for $w_0$ and $z_0$. For Gaussian beams, the beam radius $w$ at
position $z$ is given by~\cite{Kogelnik66AO}%[equation (20)]
\begin{equation}\label{Eq_wofz}
	w(z) = w_0 \sqrt{1+\left(\frac{\lambda (z-z_0)}{\pi w_0^2}\right)^2}
\end{equation}
where $w_0$ is the waist radius, $z_0$ the location of the waist and $\lambda=1064$\,nm the laser wavelength.
The \MBF{} is from a different design than \LO{} and \Tx{} fibre couplers.
In contrast to \Tx{} and \GMF{}, the intensity distributions of the \MBF{} beam showed non-Gaussian contributions.

The \GMF{} beam has a larger waist than the \MBF{} beam.
At position zero (at the \Rx{}-clip) in figure~\ref{Fig_beams} the curvatures of \MBF{} and GMF have opposite sign.
The imaging systems on the OB image the \Rx{} beam at position zero (\Rx{} clip) to the measurement QPDs.
Hence, a change in the beam curvature at the \Rx{}-clip also changes the beam curvature at
the measurement interferometer QPDs.

Table~\ref{Tab_beams}  summarises the beam parameters plotted in figure~\ref{Fig_beams}.
The table shows the waist radius and waist position along with the endpoints of the 95\% confidence intervals.
The listed waist position is  given relative to the \Rx{} clip.
The positive direction points towards the measurement interferometers.
\begin{table}
	\caption{\label{Tab_beams}Gaussian parameters of \MBF{}, \GMF{}, and \Tx{}. Here, $w_0$ and $z_0$ are the waist radius and position, respectively. 
		The indices `min' and `max' are the lower and upper endpoints of a 95\% confidence interval. All values are given in mm.}
	\begin{indented}
		\item[]\begin{tabular}{lrrrrrr}
			\br
			Beam   & $w_0$ & $w_{0,\textrm{min}}$ & $w_{0,\textrm{max}}$ & $z_0$     & $z_{0,\textrm{min}}$ & $z_{0,\textrm{max}}$ \\
			\mr
			\MBF{} & 0.817 & 0.765                & 0.876                & -1714 & -1974            & -1473         \\
			\GMF{} & 0.989 & 0.932                & 1.044                & 1594  & 1361             & 1755          \\
			\Tx{}  & 0.936 & 0.880                & 0.997                & -246  & -328             & -148          \\
			\br
		\end{tabular}
	\end{indented}
\end{table}
%%%%%%%%%%%%%%%%%%%%%%%%%%%%%%%%%%%%%%%%%%%%%%%%%%%%%%%%%%%%%%%%%%%%%%%
\subsection{Telescope simulator alignment}
\label{Sec_TSalignment}
As described in section~\ref{sec: test bed summary} the TS is aligned with respect to the OB by centring the \LO{} beam to the CQP diodes. 
For the x, y and yaw (around the z axis) degrees of freedom the TS is shifted in plane (see figure~\ref{Fig_TSonOB} for the definition of the coordinate system).
For roll, pitch, and z degree of freedom (rotations around x, y, and z, respectively) the height of the individual mounting feet 
(see figure~\ref{Fig_TSonOB} and \cite{Chwalla16CQG}) is adjusted. With this method the beam was centred on the CQP diodes with a $1-2\,\upmu$m accuracy. 
This is sufficiently accurate and was limited by beam jitter due to air movement.

%%%%%%%%%%%%%%%%%%%%%%%%%%%%%%%%%%%%%%%%%%%%%%%%%%%%%%%%%%%%%%%%%%%%%%%
\subsection{Tilt actuation of the \Rx{} beam}
\label{Sec_TiltActuation}
In section~\ref{sec: test bed summary} we described why the \Rx{} beam needs to rotate around the \Rx{} clip and the principle of how to achieve this. 
In the following, we describe the technical implementation of this concept.

The \Rx{} beam was aligned to the \LO{} beam and tilted around the centre of the \Rx{} clip with the two
piezo actuators on the TS by an automatic procedure that was implemented in the readout program.
The procedure used the position of the \Rx{} beam on the \refQPD{} on the TS and the angle
between \Rx{} and \LO{} beam at the \refQPD{}.
The \Rx{} beam position on the \refQPD{} was obtained by a power modulation of the \Rx{} beam at a
frequency of 267\,Hz imposed by the \Rx{} AOM.
Demodulation of the \refQPD{} signals in the phasemeter allowed the computation of differential
power sensing (DPS) signals~\cite{Wanner12OC}.
The calibrated DPS signals delivered the \Rx{} beam position on the \refQPD{}.
The beam angle between \Rx{} and \LO{} beams on the \refQPD{} was obtained from differential wavefront
sensing (DWS) signals~\cite{Morrison94AO, Morrison94AO2}.

The alignment procedure required that the \Rx{} beam hit the \refQPD{} under a sufficiently small
angle ($< 500$\,$\upmu$rad) so that meaningful DWS signals could be computed.
Typically, the angle between \Rx{} beam and \LO{} beam was aligned to better than 10\,$\upmu$rad and
the centring of the \Rx{} beam on the \refQPD{} was within 5\,$\upmu$m.

For a tilt-to-length coupling measurement, the \Rx{} beam was aligned in both axes.
Beam tilt was performed in the vertical axis (z-axis, see figure~\ref{Fig_TSonOB} for the definition of
the coordinate system). Starting from angle zero, beam angle steps towards negative angles, from the most
negative angle towards the most positive angle and finally towards zero were commanded.
After each angle step, the \Rx{} beam was centred on the \refQPD{}.
This ensured that the beam was tilted around the \refQPD{} and hence around the \Rx{} clip.
For each of the 74 angle steps 90\,s of data were averaged.
The mean DWS signal between \Rx{} and \LO{} beam was used as measure for the \Rx{} beam angle.
Path length signals of all diodes were recorded. In the x-axis the
\Rx{} beam was not actively actuated. During all of the 74 angle steps, the centring of the \Rx{} beam on
the \refQPD{} stayed constant to within $\pm$3\,$\upmu$m in both x and z axes.
The angle in the x axis, which was not actively controlled, stayed constant within $\pm$15\,$\upmu$rad.

%%%%%%%%%%%%%%%%%%%%%%%%%%%%%%%%%%%%%%%%%%%%%%%%%%%%%%%%%%%%%%%%%%%%%%%
\subsection{Beam angle calibration}
\label{Sec_BeamAngleCalibration}
The aim of the given experiment is to demonstrate that the TTL coupling can be reduced to less than $\pm 25\,\upmu$m/rad. 
Here, the angle given in radian is the tilt angle of the \Rx{} beam  relative to the \LO{} beam.
The DWS signal between \Rx{} and \LO{} (cf.\ phase C in table~\ref{Tab_AOMfreq}) on the \refQPD{} is correlated to this relative beam angle. 
The parameters of this correlation are initially unknown and need to be calibrated.  
As described below, this is obtained here via the DPS signal on the CQPs, for both \Rx{} beam implementations (\MBF{} and \GMF{}).

For each \Rx{} beam, its position on CQP2 was measured along with the DWS signal on the \refQPD{}
while the \Rx{} angle was varied. In the second step, \Rx{} beam angles were computed from the beam
positions, plotted as function of DWS signal and the resulting graph was fitted by a third order polynomial.
The DPS signal on CQP2 was calibrated via an analytical expression, which depends only on the diode's geometry and the spot size of the \Rx{} beam~\cite{Wanner10PhD}.
The \Rx{} spot size was taken from figure~\ref{Fig_beams} at position 601.9\,mm, which is equivalent to the CQP2 position relative to the \Rx{} clip.

%%%%%%%%%%%%%%%%%%%%%%%%%%%%%%%%%%%%%%%%%%%%%%%%%%%%%%%%%%%%%%%%%%%%%%%
\subsection{Reference interferometer longitudinal positioning}
\label{Sec_RefIfoLoPos}
As described in section~\ref{sec: test bed summary}, the reference detectors (\refSEPD{} and \refQPD{}) need to be in positions equivalent to the \Rx{} clip. 
Here, a millimetre longitudinal accuracy is sufficient~\cite{Schuster17CQG}.
This was achieved by sufficiently stringent manufacturing tolerances of the test bed and careful
positioning of the \Rx{} clip mount and the mount of the \refSEPD{}.

For the optical design, it is not the optical path length $\Sigma_i n_i \cdot d_i$ that needs to be matched but $\Sigma_i d_i/n_i$,
where $n_i$ is the refractive index and~$d_i$ the geometrical length of segment~$i$.
The quantity $\Sigma_i d_i/n_i$ is also relevant for and known from the mode propagation of Gaussian laser beams and higher order modes.

%%%%%%%%%%%%%%%%%%%%%%%%%%%%%%%%%%%%%%%%%%%%%%%%%%%%%%%%%%%%%%%%%%%%%%%
\subsection{Reference interferometer lateral alignment}
\label{Sec_RefIfoAlignment}
After the longitudinal alignment, the \refSEPD{} also needs to be aligned laterally to the \Rx{} clip.
We do this by rotating the \Rx{} beam while measuring the phase changes on the \refSEPD{} and in the \Rx{} clip, thereby comparing the TTL coupling at these two positions.
Any mismatch between these two points laterally to the \LO{} beam axis leads to linear TTL coupling while longitudinal mismatch contributes quadratically.
The differential TTL coupling in these two points was required to be less than $\pm 25\,\upmu$m/rad.

The lateral alignment was achieved in a four-step procedure.
The first step was already described in section~\ref{Sec_TSalignment} where the \LO{} beam was aligned to the CQPs. 
This established the \LO{} beam as a reference for temporary photo diodes in the \Rx{} clip on the OB.
In the second step, a QPD was placed temporarily in the \Rx{} clip and aligned to the \LO{} using DPS signals.
In the third step the temporary QPD was replaced by a temporary SEPD (identical to the \refSEPD{}) 
such that the temporary SEPD was precisely centred to the temporary QPD (within a few micrometers). 
This temporary SEPD (\tempSEPD{}) was thereby centred on the \LO{} beam.
In the fourth step, the \refSEPD{} was laterally aligned to the \tempSEPD{}.

\begin{figure}
	\centering
	\includegraphics[width=0.25\textwidth]{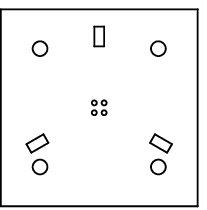}
	\hspace{10mm}
	\includegraphics[width=0.25\textwidth]{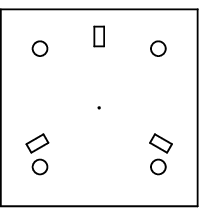}
	\caption{\label{Fig_Apertures} Apertures permanently fixed to the temporary photo diodes in the \Rx{} clip. 
		They can be attached to three spherical magnets in the \Rx{} clip via the rectangular slits. 
		Then, the centre of the SEPD is at the centre of the four holes within a few micrometers.
	}
\end{figure}

\paragraph{Alignment of the temporary diodes} The co-alignment between temporary QPD and temporary SEPD was achieved with the help of the apertures shown in figure~\ref{Fig_Apertures}. 
The left aperture is attached to the temporary QPD, the right aperture is attached to the \tempSEPD{}. The width and height of the apertures are 20\,mm. 
The central holes are the defining apertures for the QPD and the SEPD with diameters of 0.5\,mm and 0.15\,mm, respectively. 
The apertures are laser-cut from 0.1\,mm thick magnetic stainless steel foil (type AISI 430). 
They can be attached to three spherical magnets in the \Rx{} clip via the rectangular slits. 
Then, the centre of the SEPD is at the centre of the four holes within a few micrometers.

\paragraph{Lateral \refSEPD{} alignment procedure}
The lateral alignment of the \refSEPD{} has to be accurate to micrometer level and was achieved by measuring
the TTL coupling difference of the two SEPDs (\refSEPD{} and \tempSEPD{}).

The following alignment procedure was used
\begin{enumerate}
	\item Measure the TTL coupling between \Rx{} and \LO{} on the \refSEPD{} and subtract this from the corresponding TTL coupling on the \tempSEPD{}.
	\item If there is a difference move the \refSEPD{} laterally.
	\item Perform another TTL measurement.
	\item Repeat until the difference between the two SEPDs is minimized.
\end{enumerate}

\begin{figure}
	\includegraphics[width=0.5\textwidth]{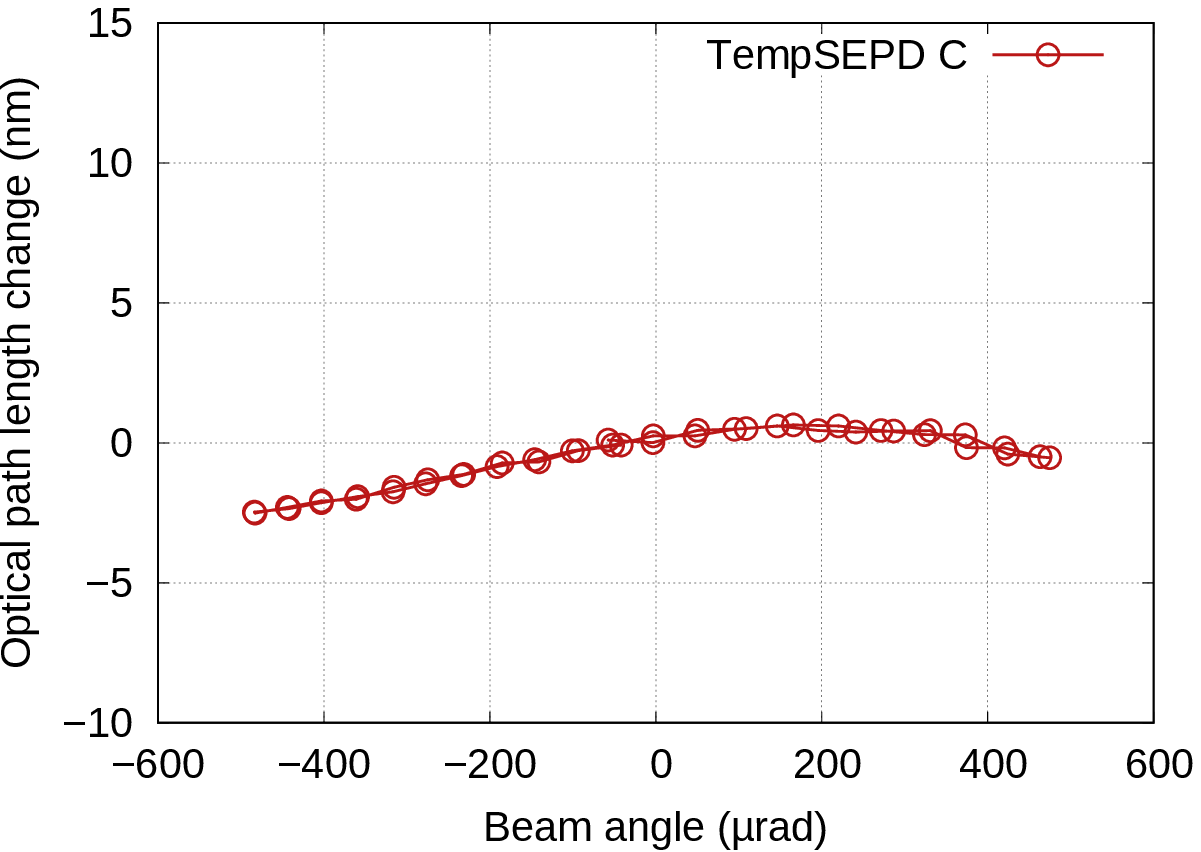}%
	\includegraphics[width=0.5\textwidth]{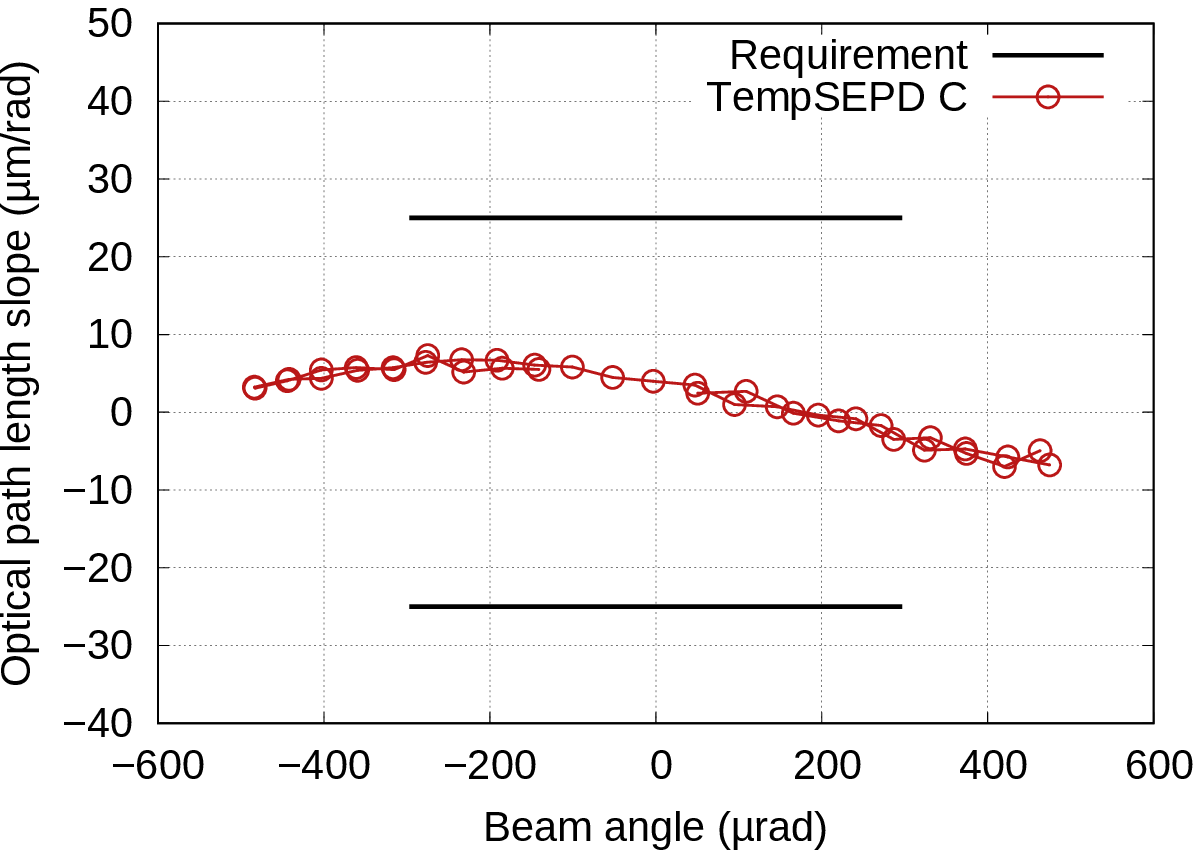}%
	\caption{Difference of the path length signal between the \refSEPD{} and the \tempSEPD{} in the \Rx{} clip with the \MBF{} beam. 
		The \refSEPD{} is aligned to minimize the path length change. 
		Left:\ Path length change vs.\ beam angle, right:\ Slope of path length change vs.\ beam angle. \label{Fig_refSEPDalignedGauss}}
\end{figure}

Figure~\ref{Fig_refSEPDalignedGauss} shows the difference of the path length signal between the \refSEPD{} 
and the \tempSEPD{} in the \Rx{} clip with the \MBF{} beam after completion of the alignment procedure for the \MBF{} beam.
The linear coupling could be reduced to be well below the requirement of 25\,$\upmu$m/rad by
aligning the \refSEPD{} laterally.
The quadratic coupling that is still visible after the alignment cannot be reduced by a lateral
alignment of the \refSEPD{}~\cite{Schuster17CQG} but is sufficiently small to fulfil the requirement.
%

%%%%%%%%%%%%%%%%%%%%%%%%%%%%%%%%%%%%%%%%%%%%%%%%%%%%%%%%%%%%%%%%%%%%%%%
\subsection{Mitigation of temperature-driven length changes}
\label{Sec_Mitigation}
The tip-tilt mount used for TS alignment was designed to be picometre-stable if required.
For this reason the feet of the mount could be clamped to the telescope simulator and the vertical alignment
screws could be retracted. The TS then had a connection to the OB entirely of low-expansion material
(Zerodur by Schott for the TS baseplate and connection blocks and Ultra-Low
Expansion (ULE) glass by Corning for the feet).

\begin{figure}
	\centering
	\includegraphics[width=0.5\textwidth]{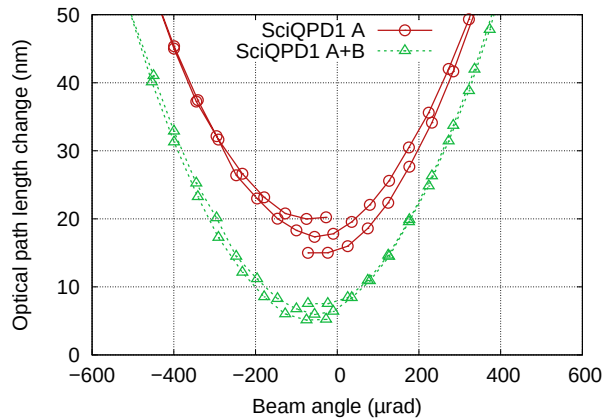}
	\caption{\label{Fig_A+B_vs_A}Path length change between \SciQPD{}1 and \refSEPD{} plotted over the beam angle for a comparison between A phase and A+B phase.  
		The traces have been vertically shifted for clarity. The A phase signal shows a drift that is removed in the A+B phase signal.
	}
\end{figure}

In our experiment we omitted clamping of the feet and retraction of the alignment screws because we did not require picometre stability and it allowed easier handling of the TS.
Consequently, in some TTL measurements we observed temperature-driven drifts of the path length signals
(see figure~\ref{Fig_A+B_vs_A} and figure~9 in reference~\cite{Chwalla16CQG}).
We believe these drifts to be dominated by the thermally-driven expansion of the alignment screws.
We used a combination of measurement signals to remove this effect.

The height of the TS is measured in two phase signals: in signal A between \Tx{} and \Rx{} and in signal B
between \LO{} and \Tx{}. Since the phase B measures the phase relation between two stable beams, it is a
good measurement of the height variation of the TS.
In the following we will use the B phase signal to correct for the height variations of the TS in the A phase.
Thus we minimise our sensitivity to TS movement in the z direction.

Figure~\ref{Fig_A+B_vs_A} shows a comparison between only A phase and A+B phase when no imaging
system was present in the measurement interferometer. In the only A phase trace a drift can be observed
which is caused by a height change of the TS. If the B phase is added, the drift disappears and the
measured curve is not affected by the height variations any more.

In figure~\ref{Fig_A+B_vs_A} and all following figures that use QPDs the phase signals of the four segments
of each QPD are averaged (see equation~(5) in~\cite{Wanner15JPCS}).
From this average the phase signal of
the \refSEPD{} is subtracted. The phase difference $\Delta\phi$ is converted to an optical length change
$\Delta s$ according to $\Delta s=\frac{\lambda}{2\pi}\Delta \phi$.

%%%%%%%%%%%%%%%%%%%%%%%%%%%%%%%%%%%%%%%%%%%%%%%%%%%%%%%%%%%%%%%%%%%%%%%
\subsection{Alignment of photo diodes and imaging systems}
Before the two imaging systems can be placed on the optical bench in their nominal position, they need to be assembled and pre-aligned.
Therefore, the lenses (and lens pairs in the case of the four-lens imaging system) are held by adjustable mounts, fixed to a ``super baseplate'' which can be attached to the OB.

For the pre-alignment of the two-lens imaging systems we used a QPD and a tiltable beam.
We performed the alignment in two steps. In the first step lens~2 and later lens~1 were aligned on a centre
beam at their nominal longitudinal positions. In the second alignment step a tilting beam was used.
The beam walk on the QPD behind the imaging system was minimized by changing the distance
between the lenses.

For the pre-alignment of the four-lens imaging systems we used a Shack-Hartmann sensor (\SHS{}) with
built-in light source and a double-pass configuration.
In the alignment steps, we placed the imaging system in between \SHS{} and a plane mirror and
minimised wavefront errors. First, we aligned the lens pair L1+L2, removed it from the super baseplate,
rotated the super baseplate by 180$^{\circ}$ and inserted and aligned lens pair L3+L4.
Then we aligned the lens pairs to each other on the super baseplate.

After these steps, the imaging systems are pre-aligned and then need to be placed on the OB such that the \Rx{} clip is imaged to the \SciQPD{}s.
That means, the \SciQPD{}s sense no beam walk when the \Rx{} beam is rotated around the \Rx{} clip. In a classical imaging system the \SciQPD{}s 
are thereby located in the exit pupil, while the \Rx{} clip defines the entrance pupil.

\subsection{Optimisation of the diode position}
\label{sec: photodiode alignment}
Placing the \SciQPD{}s into the exit pupil does not necessarily define minimal TTL coupling.
The point to point imaging ensures that there is no geometrical TTL coupling. 
That means, assuming a perfect imaging system and plane waves without clipping there is no TTL coupling on the \SciQPD{}s. 
However, the use of Gaussian laser beams, imperfect imaging, and clipping on the QPDs' slits and outer borders result in a 
residual TTL coupling on the \SciQPD{}s located in the exit pupil. As mentioned before and described in detail in~\cite{Schuster17CQG}, 
lateral (longitudinal) shifts of the diodes  result in additional  linear (quadratic) TTL coupling.
Therefore, we shift the \SciQPD{}s intentionally laterally to minimize any residual linear TTL coupling and longitudinally to reduce any residual quadratic TTL coupling.
Effectively we thereby counteract the TTL coupling of the described non geometric sources with two geometric effects.
In this work, we call the position with minimal TTL coupling ``optimal position''.
All measurements shown in section~\ref{Sec_Results} are performed with the \SciQPD{}s in optimal position.

%%%%%%%%%%%%%%%%%%%%%%%%%%%%%%%%%%%%%%%%%%%%%%%%%%%%%%%%%%%%%%%%%%%%%%%
\section{Results}
\label{Sec_Results}
\subsection{Tilt-to-length coupling using the \GMF{} beam}
\begin{figure}
	\includegraphics[width=0.5\textwidth]{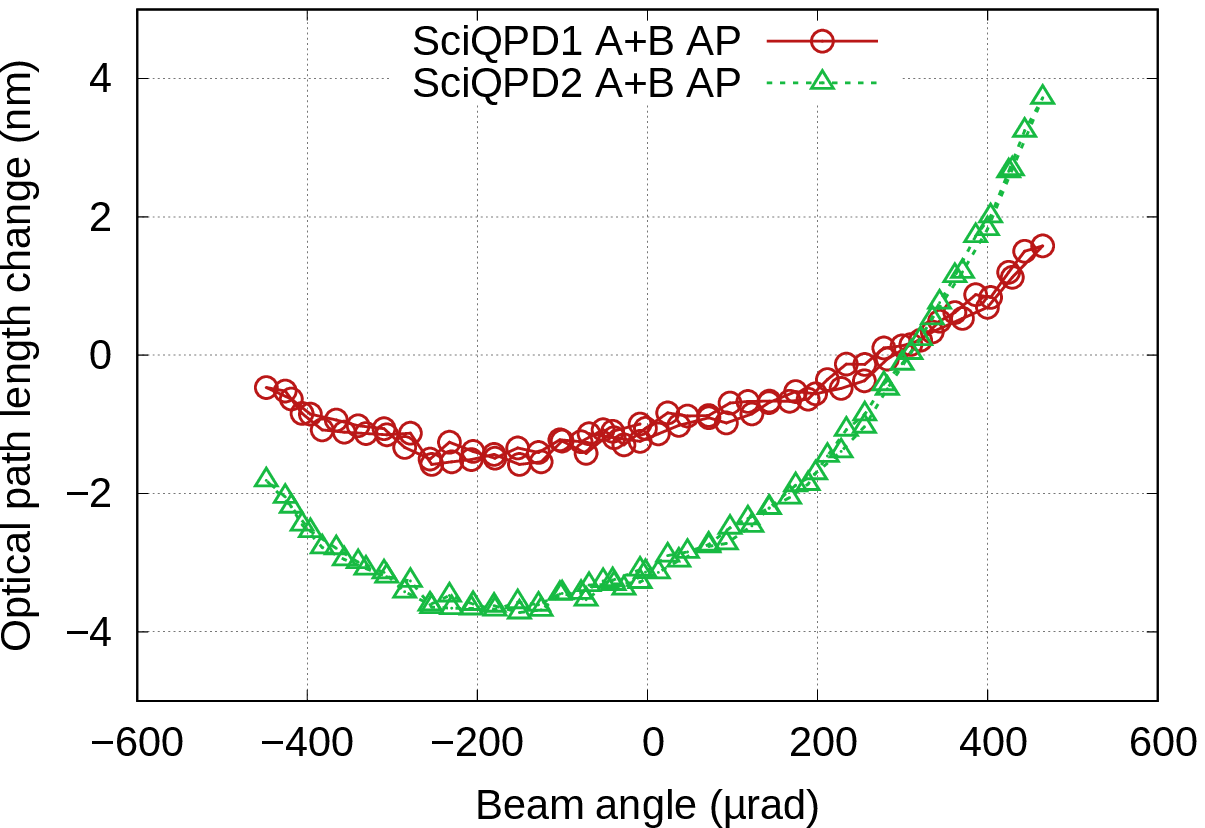}%
	\includegraphics[width=0.5\textwidth]{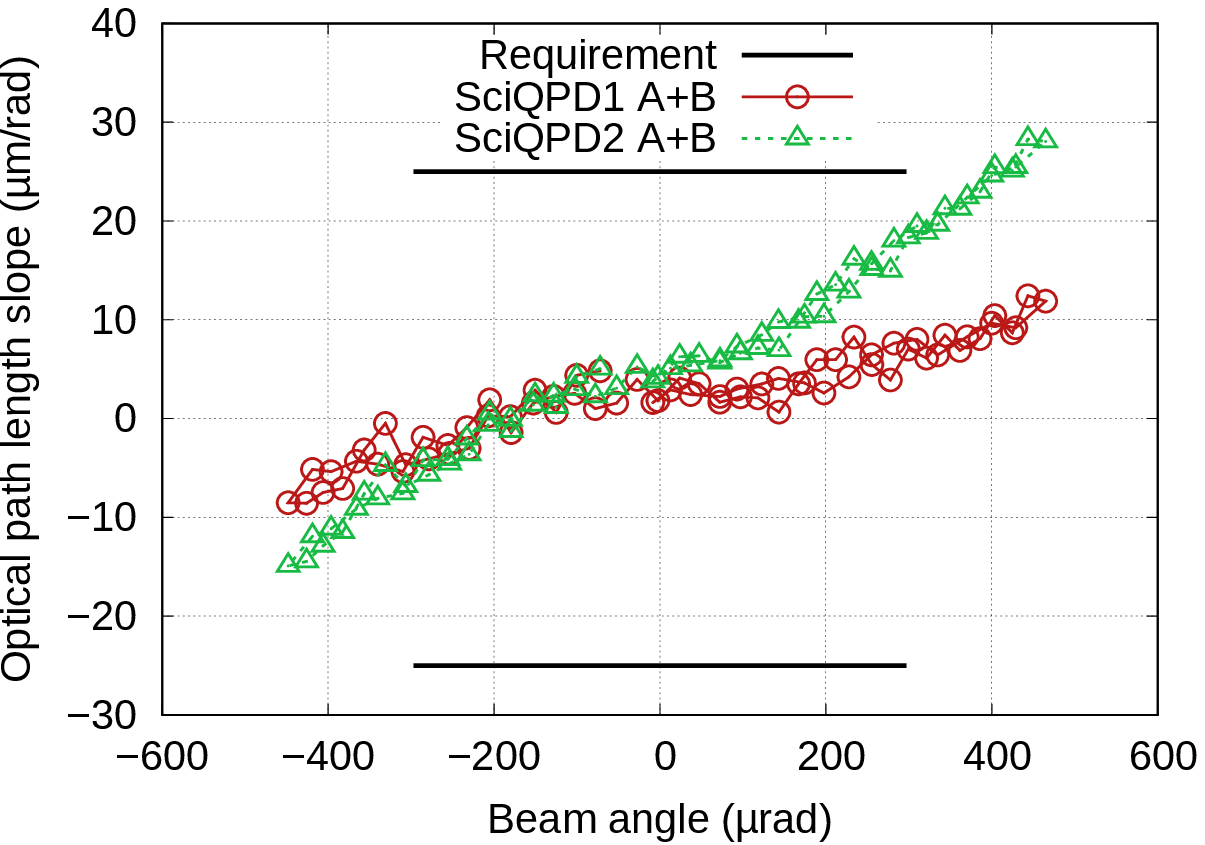}%
	\caption{\label{Fig_GMF2-2Lens}Two-lens imaging systems performance using the \GMF{} beam; 
			Both two-lens imaging systems fulfil the TTL requirement.}
\end{figure}

\begin{figure}
	\includegraphics[width=0.5\textwidth]{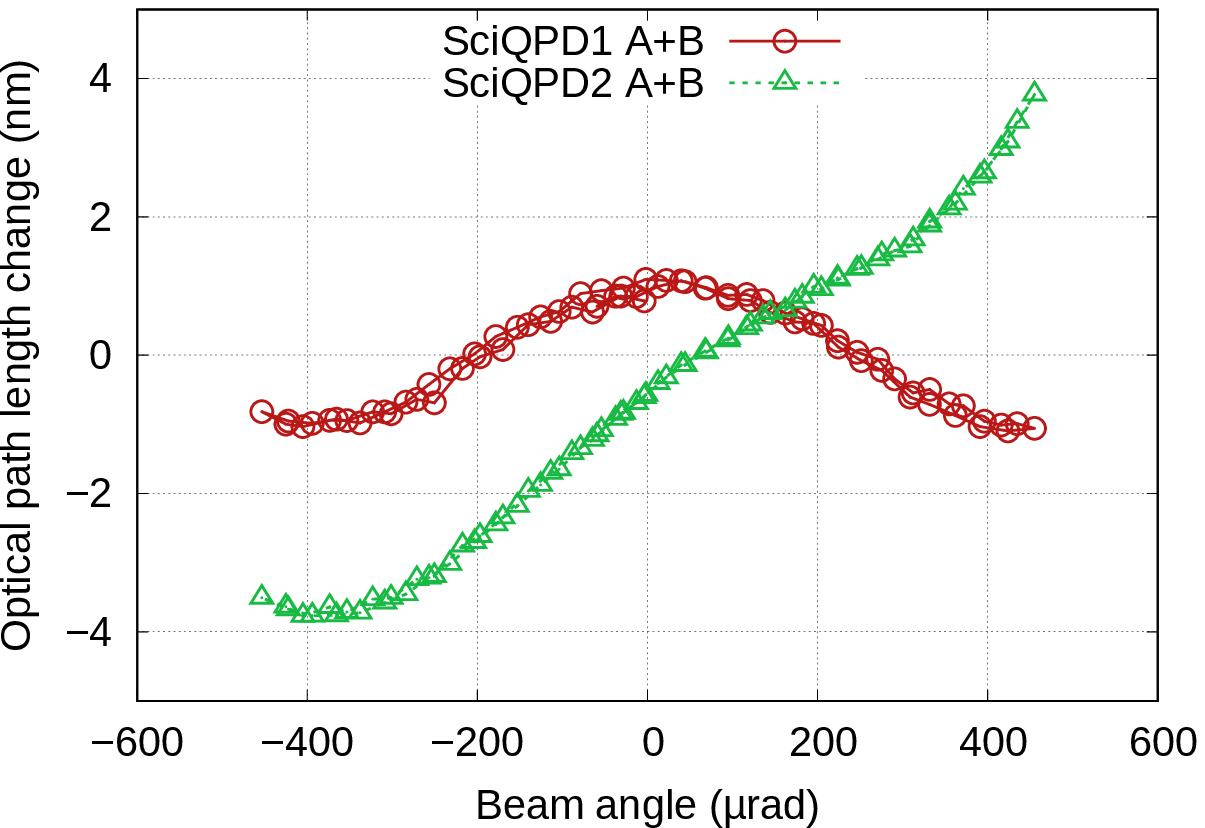}%
	\includegraphics[width=0.5\textwidth]{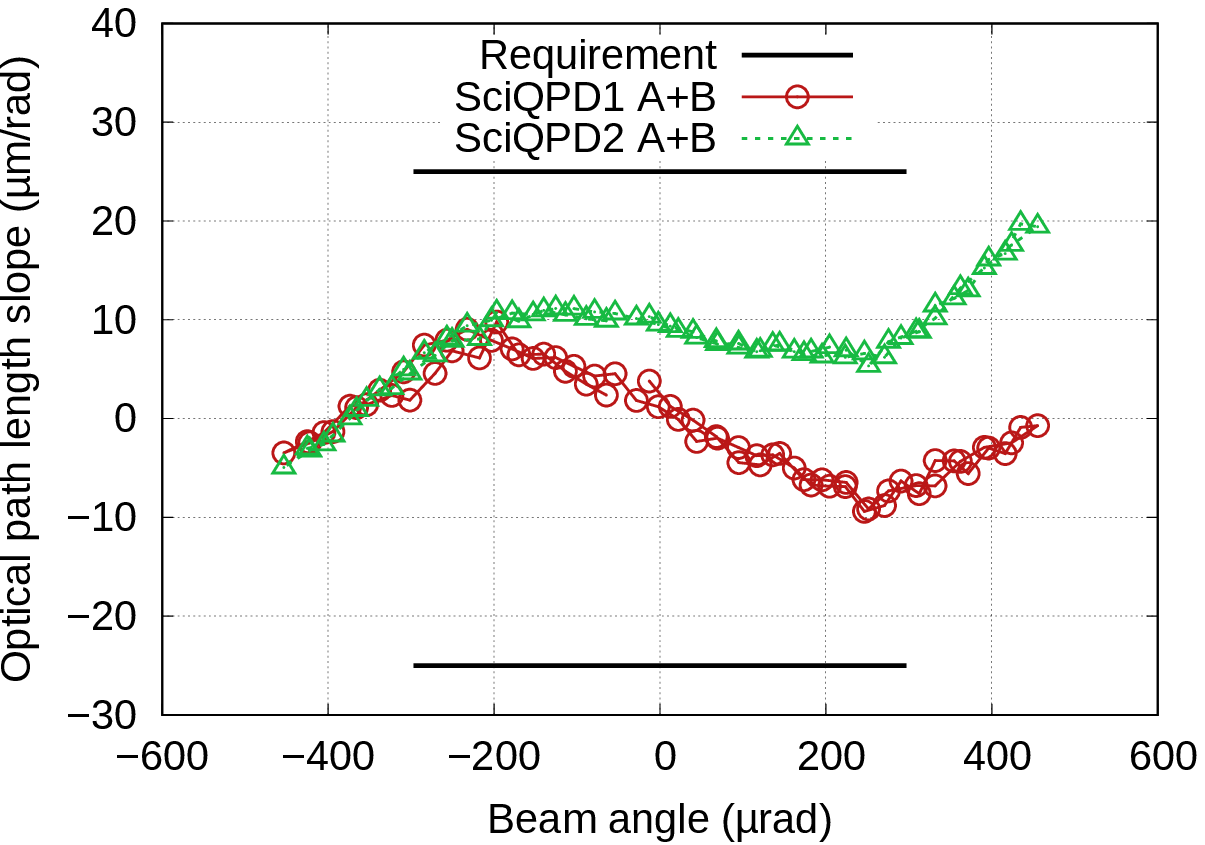}%
	\caption{\label{Fig_GMF2-4Lens}Four-lens imaging systems performance using the \GMF{} beam; 
		Both four-lens imaging systems fulfil the TTL requirement.}
\end{figure}

Figures~\ref{Fig_GMF2-2Lens} and \ref{Fig_GMF2-4Lens} show the TTL coupling using the \GMF{} behind the two-lens
and four-lens imaging systems, respectively.
The left side shows path length change vs.\ beam angle, the right side shows the slope of
path length change vs.\ beam angle.
For both \SciQPD{}s and both imaging systems the path length slopes are well within
the requirement.
Since our aim was to show that TTL coupling in the measurement interferometer could be brought within the
requirement using adequate imaging systems, no attempts were made to further reduce the TTL
coupling or match the coupling of the two \SciQPD{}s.

%%%%%%%%%%%%%%%%%%%%%%%%%%%%%%%%%%%%%%%%%%%%%%%%%%%%%%%%%%%%%%%%%%%%%%%
\subsection{Tilt-to-length coupling using the \MBF{} beam}
\begin{figure}
	\includegraphics[width=0.5\textwidth]{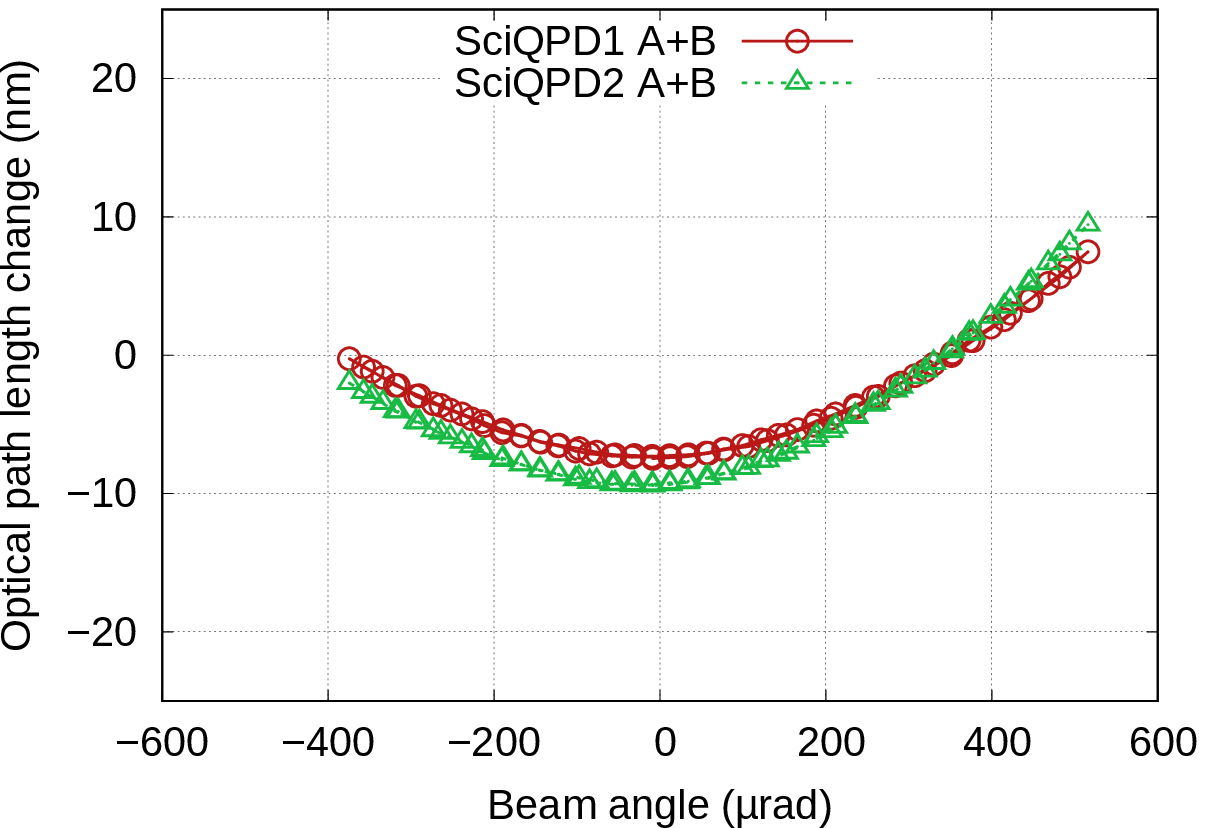}%
	\includegraphics[width=0.5\textwidth]{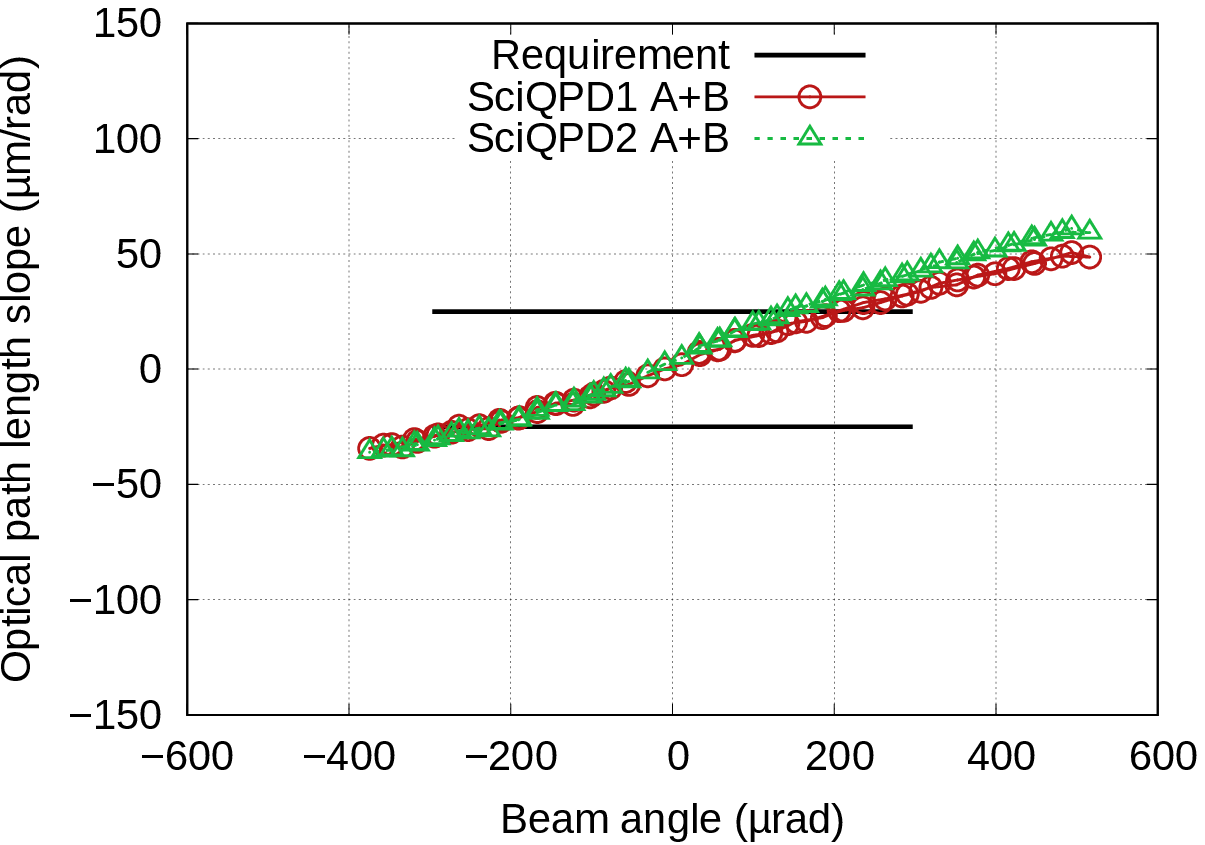}%
	\caption{Two-lens imaging systems performance using the \MBF{} beam;
		Changing the \Rx{} beam (from the \GMF{}) to the \MBF{} leads to a violation of the requirement for both two-lens imaging systems.\label{Fig_RXG-2Lens} }
\end{figure}

\begin{figure}
	\includegraphics[width=0.5\textwidth]{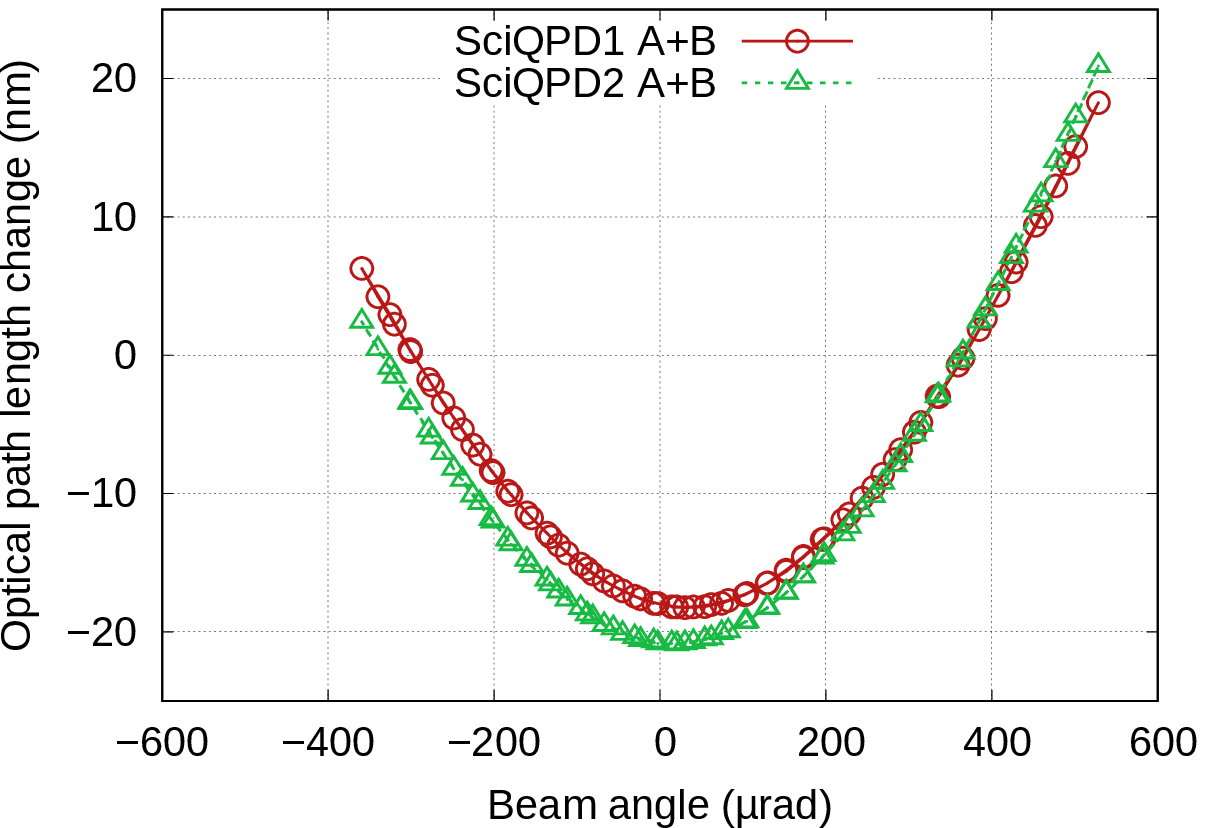}%
	\includegraphics[width=0.5\textwidth]{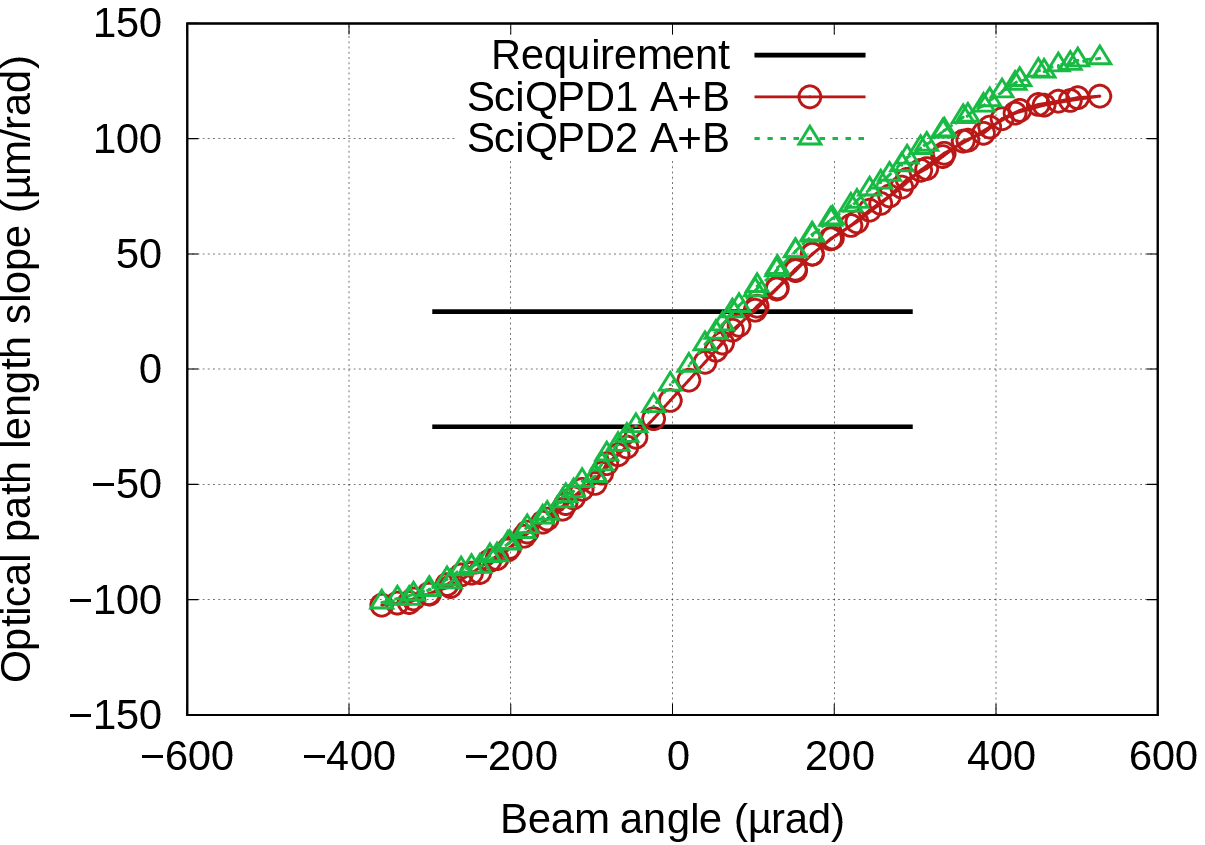}%
	\caption{Four-lens imaging systems performance using the \MBF{} beam;
		Changing the \Rx{} beam (from the \GMF{}) to the \MBF{} leads to a violation of the requirement for both four-lens imaging systems. \label{Fig_RXG-4Lens} }
\end{figure}

Figure~\ref{Fig_RXG-2Lens} shows the TTL coupling behind the two-lens imaging systems using the \MBF{}.
For an \Rx{} beam tilt of $\pm 300\,\upmu$rad a residual coupling of -30\,$\upmu$m/rad to 40\,$\upmu$m/rad was obtained, which is violating the requirement of $\pm$25\,$\upmu$m/rad.
The same measurement with the four-lens imaging systems using the \MBF{} beam is shown in
figure~\ref{Fig_RXG-4Lens}.
The residual TTL coupling was $\pm100\,\upmu$m/rad at $\pm 300\,\upmu$rad \Rx{} beam angle, clearly violating the requirement of $\pm$25\,$\upmu$m/rad.

\subsection{Discussion}
These measurements were performed at the optimal \SciQPD{} positions (cf.\ section~\ref{sec: photodiode alignment}), such that the shown residual TTL coupling could not be reduced further. 

We have shown, that both types of imaging systems violated the requirement of $\pm$25\,$\upmu$m/rad for \Rx{} beam tilts of $\pm 300\,\upmu$rad if the \MBF{} beam is used, 
but perform according to the requirement if the \GMF{} beam is used. 
The performance of the imaging systems therefore clearly depends on beam properties. 

The underlying mechanisms resulting in the violation of the requirement while using the \MBF{} are discussed in section~\ref{Sec_Simulations}.

%%%%%%%%%%%%%%%%%%%%%%%%%%%%%%%%%%%%%%%%%%%%%%%%%%%%%%%%%%%%%%%%%%%%%%%
\section{Tilt-to-length coupling simulations}
\label{Sec_Simulations}
In order to analyse the dependency of the TTL suppression performance on beam properties, we performed computer simulations using the software IfoCAD~\cite{Wanner12OC}.
The optical setup with the different imaging systems and the optimisation to minimise the TTL coupling by shifting the photo diode  (cf.\ section~\ref{sec: photodiode alignment}) 
was simulated and the influence of different beam parameters to the residual TTL was computed.

\subsection{Simulation algorithm}
\label{sec:Simulation algorithm}
The goal of this simulation is to show the best possible TTL coupling behind an imaging system
(best possible:\ optimal  QPD position for minimal coupling) as a function of the beam
parameters --  \Rx{} and \Tx{} waist radius and position -- to find correlations between the different properties.
In detail, the simulation follows the algorithm shown here:

\begin{enumerate}
	\item The simulation assumes a perfectly aligned imaging system (either the two-lens or the four-lens system),
	      the point of rotation is at position zero and all lenses and the QPD are at their nominal positions.
	\item The TTL coupling for given beam parameters of the \Tx{} and \Rx{} beam is computed in the range
	      of $0\,\upmu$rad to $300\,\upmu$rad. The maximum in the optical path length slope in this range is called $c_\mathrm{TTL}$. 
	      The simulated setup is spherically symmetric so an extension of the angular
	      range down to $-300\,\upmu$rad will produce identical results.
	\item The QPD is then placed in its optimal position, i.e.\ it is shifted longitudinally until the TTL coupling ($c_\mathrm{TTL}$) is minimised. 
	The optimal position is labeled $d_\mathrm{QPD}$ and the TTL coupling at this optimised QPD location is called $c_\mathrm{opt}$.
	\item Next, the beam parameters of the \Rx{} beam (waist radius and waist position) are varied and the corresponding
	      optimised couplings $c_\mathrm{opt}$ and the optimal QPD positions $d_\mathrm{QPD}$ are computed
	      for all combinations in the range of $\omega_0 = 0.7\,$mm to $1.1\,$mm and $z_0 = -2\,$m to $2\,$m.
	      The results are plotted in a heat map, where the x axis is the waist radius of the \Rx{} beam, the y axis is
	      the waist position of the \Rx{} beam and the colour identifies the optimal coupling
	      $c_\mathrm{opt}$ in $\upmu$m/rad (e.g.\ one sub-plot of figure~\ref{fig: two lens slope multi}).
	\item The steps (i) to (iv) are repeated for different sets of \Tx{} beam parameters, resulting in the shown multi-plots 
	(e.g.\ figure~\ref{fig: two lens slope multi}). Here, the beam parameters of the \Tx{} beam were chosen to cover the parameter range of the beams present in the given test-bed (c.f.\ table~\ref{Tab_beams}.)
	\item Finally, coloured crosses are placed in each sub-plot, illustrating the fitted parameters and confidence intervals for
the \MBF{} (green) and the \GMF{} (purple) beam.
	\end{enumerate}
The simulations therefore did not include imperfections of the imaging systems, beam imperfections such as deviations from the fundamental Gaussian model, 
or phase changes from the misalignment of the \refSEPD{}. Furthermore, the optimisation of the photo diodes alignment was stopped in the experiment once the 
TTL couplings were within the requirement, which is not reflected in the given simulation. Accordingly, a match of the computed TTL coupling factors with 
those obtained experimentally is not envisaged. Instead, the simulation fully focuses on the impact of beam parameters to the resulting TTL coupling.

\subsection{Results for the four-lens imaging system}
\label{sec: multiplot 4l}
\begin{figure}
	\centering
	\includegraphics[width=0.7\textwidth]{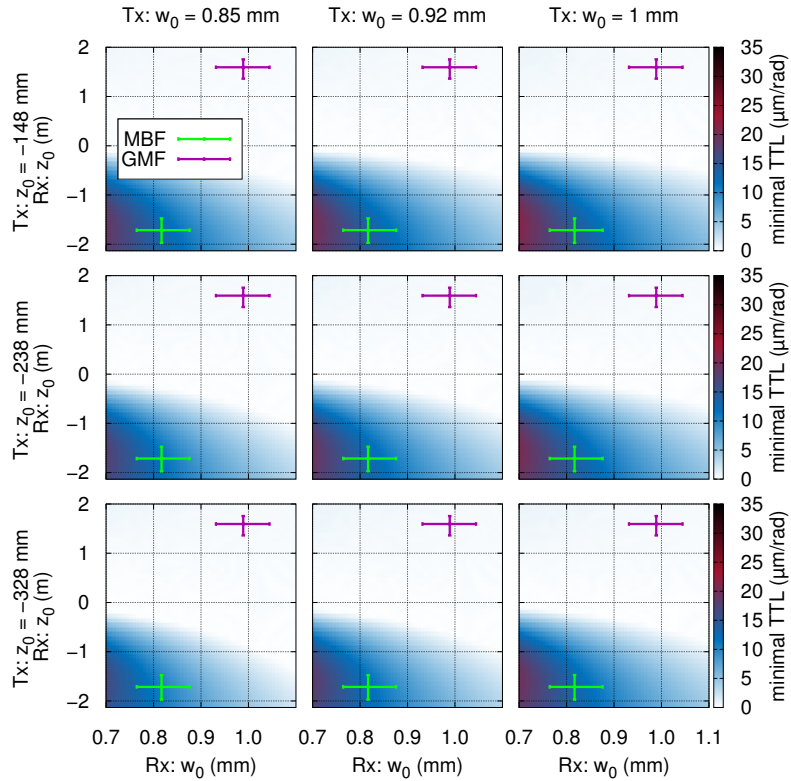}
	\caption {Simulated TTL coupling $c_\mathrm{opt}$ of the four-lens imaging system obtained following the 
		procedure described in section~\ref{sec:Simulation algorithm}. This shows a significant TTL coupling for the \MBF{} beam, 
		while the TTL coupling using the \GMF{} beam is negligible.}
	\label{fig: four lens slope multi}
\end{figure}

Figure~\ref{fig: four lens slope multi}  shows the multi-plot for the TTL slope $c_\mathrm{opt}$ behind the four-lens imaging system.

For the \GMF{} the optimal TTL coupling is negligible and far below the requirement. In contrast, for the \MBF{} the residual 
TTL coupling is in the order $10-20\,\upmu$m/rad. Thereby, the beam parameter dependency observed in the experiment is clearly reflected in this simulation. 

\begin{figure}
	\centering
	\includegraphics[width=0.7\textwidth]{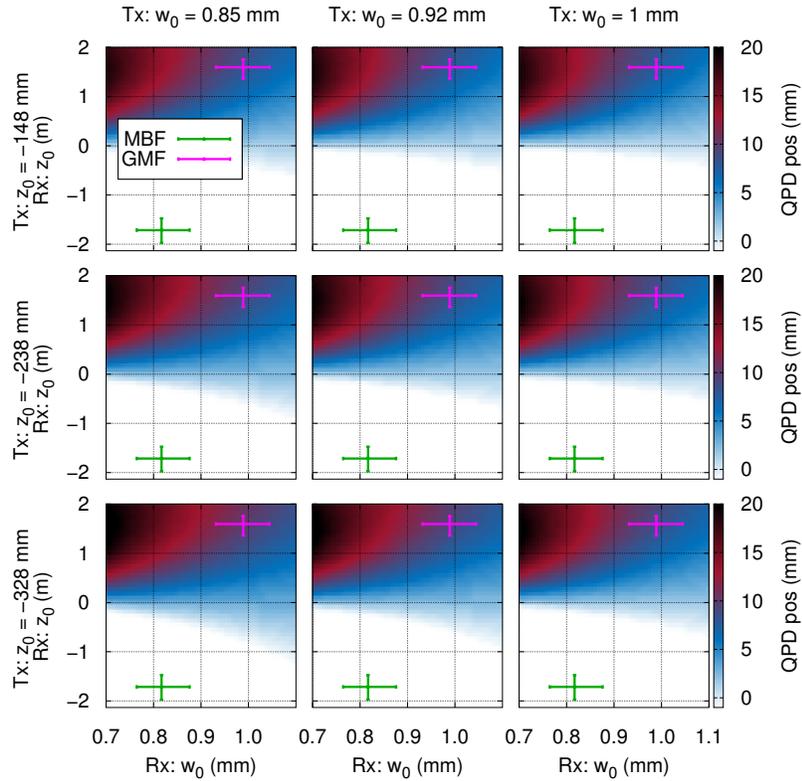}
	\caption{Optimal QPD position of the four-lens imaging system, corresponding to figure~\ref{fig: four lens slope multi}. 
		This shows that when the \MBF{} beam is used, the QPD is shifted by -1\,mm, resulting in the closest possible distance behind the imaging system.}
	\label{fig: four lens QPD pos multi}
\end{figure}

Figure~\ref{fig: four lens QPD pos multi} shows the corresponding computed optimal QPD positions.
For the \GMF{} beam, this optimal QPD position is about 10\,mm, that means the distance between imaging system and QPD was 
increased by about 10\,mm with respect to the nominal position, resulting in the low TTL coupling shown in figure~\ref{fig: four lens slope multi}. 
For the \MBF{} beam the optimal position was computed to be -1\,mm. This value was defined in the simulation to be the closest position to the imaging 
system allowed for the QPD, limited by the mechanical lens mount. 
In theory, the coupling could be reduced further by shifting the QPD closer to the imaging system or even virtually into it. 
This shows, that for a full compensation of the non-geometrical coupling effects, 
the QPD would need to be shifted closer to the imaging system, which however is not possible due to geometrical constraints. 
Due to these constraints, the compensation was incomplete, resulting in the high TTL coupling in figure~\ref{fig: four lens slope multi} and the observed violation of the requirement in the experiment.

\subsection{Results for the two-lens imaging system}
\label{sec: multiplot 2l}
\begin{figure}
	\centering
	\includegraphics[width=0.7\textwidth]{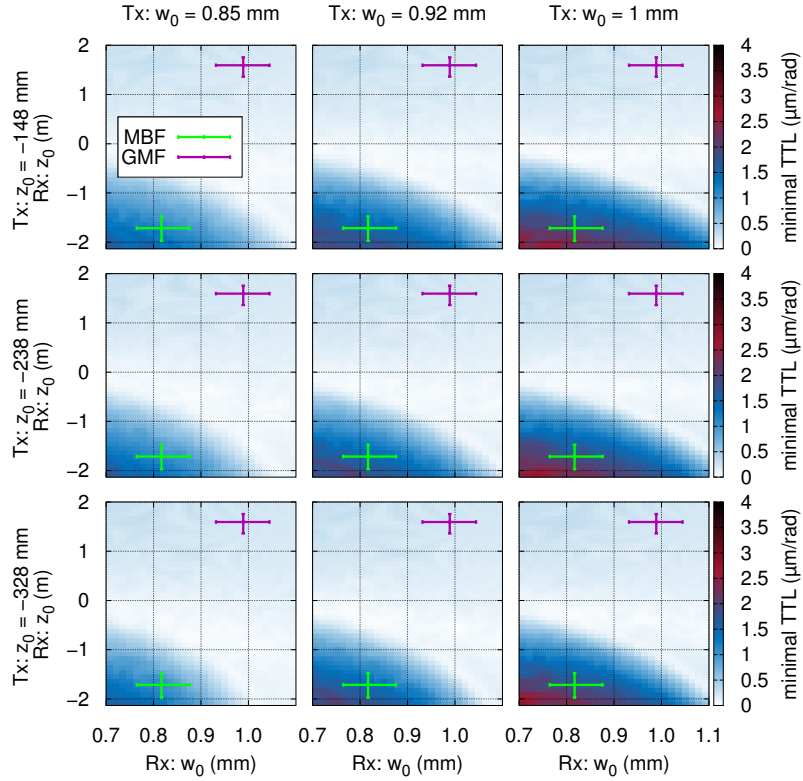}
	\caption {Simulated TTL coupling $c_\mathrm{opt}$ of the two-lens imaging system. Again, a TTL coupling for the \MBF{} beam can be observed, while this coupling is negligible if the \GMF{} beam is used.}
	\label{fig: two lens slope multi}
\end{figure}

Figure~\ref{fig: two lens slope multi} shows the multi-plot for the  two-lens imaging system.
Again, the TTL coupling is close to zero for the \GMF{} beam and with about $3\,\upmu$m/rad significantly higher for the \MBF{} beam. 
This value, however, is considerably smaller than observed in the corresponding experiment. 
The most likely explanation for this deviation is the focussing of the non-Gaussian contributions (mentioned in section~\ref{sec:Beams}) of the \MBF{} beam in the two-lens imaging system.

\begin{figure}
	\centering
	\includegraphics[width=0.7\textwidth]{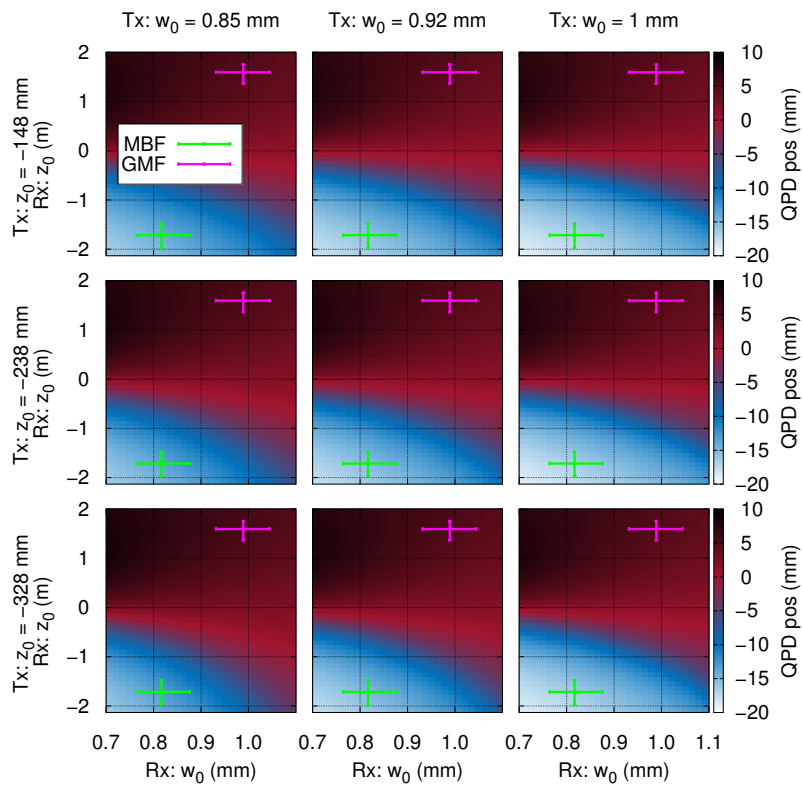}
	\caption {Optimal QPD position of the two-lens imaging system, corresponding to figure~\ref{fig: two lens slope multi}. The QPD positions ranging between -20\, to +10\,mm are not restricted by the experimental setup.}
	\label{fig: two lens QPD pos multi}
\end{figure}

In Figure~\ref{fig: two lens QPD pos multi} the optimal QPD position for the two-lens imaging system is
shown, again as a function of the \Tx{} and \Rx{} parameters.

For the two-lens imaging system the reason for this behaviour is not the limited QPD shifting range as for the
four-lens imaging systems. All optimal photo diode positions are well within the available range.

\begin{figure}
	\centering
	\includegraphics[width=0.7\textwidth]{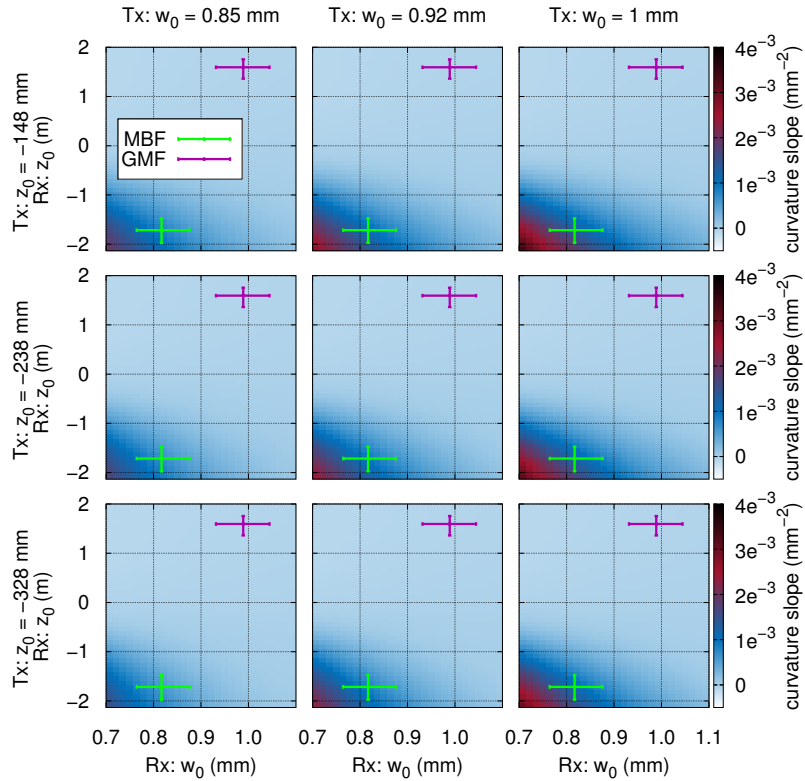}
	\caption{Simulated beam curvature difference slope at the optimal QPD position behind the two-lens imaging system,  corresponding to figure~\ref{fig: two lens slope multi}. 
		A clear correlation between the TTL coupling and the curvature difference slope can be observed.}
	\label{fig: two lens curv slope multi}
\end{figure}

We know, that the TTL coupling depends on the difference of the wavefront curvature of the interfering beams. 
In the case of the four-lens system, the output beams are collimated (defined here by a maximal Rayleigh range). 
That means the curvature difference does not change over the available QPD shifting range. 
The two-lens system however, generates diverging output beams, such that the curvature difference depends on the QPD position. The curvature difference slope $\rho_\mathrm{s}$  defined as 
\begin{equation}
	\rho_\mathrm{s}=\frac{\mathrm{d}(\rho_\mathrm{\Rx} - \rho_\mathrm{\Tx})}{\mathrm{d}d_\mathrm{tmp}}\bigg|_{d_\mathrm{tmp}=d_\mathrm{QPD}}\,,
\end{equation}
describes this position dependency, 
with $\rho_\mathrm{\Rx}$ and $\rho_\mathrm{\Tx}$ being the curvatures of the \Rx{} and
the \Tx{} wavefronts, $d_\mathrm{tmp}$ is the longitudinal position of the QPD
and $d_\mathrm{QPD}$ is the optimal position of the QPD with minimal TTL coupling.
This curvature difference is plotted in figure~\ref{fig: two lens curv slope multi} and shows a clear correlation to the TTL coupling plotted in figure~\ref{fig: two lens slope multi}.
This correlation and the non-negligible TTL coupling in the two-lens imaging system using the \MBF{} beam can probably be explained as follows. 
As described in section~\ref{sec: photodiode alignment}, the aim is to null the quadratic 
TTL coupling originating from various non-geometric effects by an additional quadratic coupling induced by longitudinally shifting the QPD.
While shifting the QPD, a curvature slope generates an additional quadratic coupling. 
With the beam parameters of the \MBF{} the curvature slope coupling counteracts the coupling induced by the QPD shift. 
Consequently, is not possible to null the quadratic TTL coupling from other non-geometric effects by changing the longitudinal position of the photo diode.

%%%%%%%%%%%%%%%%%%%%%%%%%%%%%%%%%%%%%%%%%%%%%%%%%%%%%%%%%%%%%%%%%%%%%%%
\section{Conclusion}
We have demonstrated the use of imaging systems to minimise tilt-to-length (TTL)
coupling in a setup representative for the LISA test mass interferometer. 
That means, we have shown experimentally for a two-lens and a four-lens imaging system that the TTL coupling could be reduced below the required $\pm$25\,$\upmu$m/rad for angles
within $\pm$300\,$\upmu$rad between the interfering beams. 
We observed a beam parameter dependent performance of the imaging systems. This dependency was modelled numerically and explained. 
This shows additionally, that the TTL suppression of imaging systems can be modelled and predicted by numerical simulations.

\section*{Acknowledgements}
We acknowledge funding by the European Space Agency within the project Optical Bench Development for
LISA (22331/09/NL/HB), support from UK Space Agency, University of Glasgow, Scottish Universities Physics
Alliance (SUPA), and support by Deutsches Zentrum f\"ur Luft und Raumfahrt (DLR) with funding from  the
Bundesministerium f\"ur Wirtschaft und Technologie (DLR project reference 50 OQ 0601).
We thank the German Research
Foundation for funding the cluster of Excellence QUEST - Centre for Quantum Engineering and
Space-Time Research.
Advanced imaging systems for future gravity missions were investigated in the frame of SFB1128 geo-Q and the dependency on beam parameters shown here was found. 
The simulations described in section~\ref{Sec_Simulations} show results found within geo-Q project A05, 
adapted to the LISA imaging systems. We therefore gratefully acknowledge Deutsche Forschungsgemeinschaft (DFG) for funding geo-Q.

\section*{References}
\bibliography{References}
\end{document}